\documentclass[preprint,nofootinbib]{revtex4}
  
  \usepackage{amssymb,amsmath,graphicx,color}
  \usepackage[colorlinks]{hyperref}
  \begin{document}

  %%%%%%%%%%%%%%%%%%%%%%%%%%%%%%%%%%%%%%%%%%%%%%%%%%%%%%%%%%%
  \title{$B_{u}$ ${\to}$ ${\psi}M$ decays and $S$-$D$ wave mixing effects}
  \author{Yueling Yang}
  \affiliation{Institute of Particle and Nuclear Physics,
              Henan Normal University, Xinxiang 453007, China}
  \author{Yupei Guo}
  \affiliation{Institute of Particle and Nuclear Physics,
              Henan Normal University, Xinxiang 453007, China}
  \author{Junfeng Sun}
  \affiliation{Institute of Particle and Nuclear Physics,
              Henan Normal University, Xinxiang 453007, China}
  \author{Na Wang}
  \affiliation{Institute of Particle and Nuclear Physics,
              Henan Normal University, Xinxiang 453007, China}
  \author{Qin Chang}
  \affiliation{Institute of Particle and Nuclear Physics,
              Henan Normal University, Xinxiang 453007, China}
  \author{Gongru Lu}
  \affiliation{Institute of Particle and Nuclear Physics,
              Henan Normal University, Xinxiang 453007, China}

  %%%%%%%%%%%%%%%%%%%%%%%%%%%%%%%%%%%%%%%%%%%%%%%%%%%%%%%%%%%
  \begin{abstract}
  The $B_{u}$ ${\to}$ ${\psi}M$ decays are studied with the perturbative
  QCD approach, where the psion ${\psi}$ $=$ ${\psi}(2S)$, ${\psi}(3770)$,
  ${\psi}(4040)$ and ${\psi}(4160)$, and the light meson $M$ $=$ ${\pi}$,
  $K$, ${\rho}$ and $K^{\ast}$. The factorizable and nonfactorizable
  contributions, and the $S$-$D$ wave mixing effects on the psions
  are considered in the calculation.
  With appropriate inputs, the branching ratios for the $B_{u}$ ${\to}$
  ${\psi}K$ decays are generally coincident with the experimental data
  within errors. However, due to the large theoretical and
  experimental errors, it is impossible for the moment to give a
  severe constraint on the $S$-$D$ wave mixing angles.
  \end{abstract}
  \maketitle
  %%%%%%%%%%%%%%%%%%%%%%%%%%%%%%%%%%%%%%%%%%%%%%%%%%%%%%%%%%%
  \section{Introduction}
  \label{sec01}

  The exclusive $B$ meson decays into one psion (${\psi}$) and one
  light meson ($M$) are of great interest, and have attracted much
  attention over the past years.
  In this paper, unless otherwise specified, the symbol ${\psi}$ denotes
  the high excited charmonium states with the quantum number\footnotemark[1] $I^{G}J^{PC}$
  $=$ $0^{-}1^{--}$, including ${\psi}(2S)$\footnotemark[2],
  ${\psi}(3770)$\footnotemark[3], ${\psi}(4040)$\footnotemark[3],
  and ${\psi}(4160)$\footnotemark[3] \cite{pdg};
  and the symbol $M$ refers to the members
  of the ground $SU(3)$ pseudoscalar $P$ and vector $V$ meson nonet,
  $P$ $=$ ${\pi}$ and $K$; $V$ $=$ ${\rho}$ and $K^{\ast}$.
  From the theoretical point of view, the $B$ ${\to}$ ${\psi}M$ decays are
  predominantly induced by the process $b$ ${\to}$ $c$ $+$ $W^{{\ast}-}$ ${\to}$
  $c$ $+$ $\bar{c}\,q$ ($q$ $=$ $d$ or $s$) with the spectator quark ansatz.
  The $c$ quark originating from the $b$ quark decay must unite with
  the $\bar{c}$ quark arising from the virtual $W^{{\ast}-}$ decay to
  form the flavor-singlet psion. In addition, the color charges
  of the $c$ and $\bar{c}$ quarks from two different sources must match
  with each other to be colorless.
  Hence, the $B$ ${\to}$ ${\psi}M$ decays induced by the internal $W$-emission
  interactions are color suppressed (class-II), in comparison with the
  nonleptonic $B$ weak decays induced by the external $W$-emission
  interactions (class-I).
  \footnotetext[1]{The symbols
  of $I$, $J$, $G$, $P$, $C$ refer to the isospin, angular momentum,
  $G$-parity, $P$-parity, and $C$-parity of one particle, respectively.}
  \footnotetext[2]{The symbols $nL$ in parentheses are the radial quantum number
  $n$ and the orbital angular momentum $L$,
  with $n$ $=$ $1$, $2$, ${\cdots}$, and $L$ $=$ $S$, $P$, $D$, ${\cdots}$.
  The ${\psi}(2S)$ particle is thought to be a $2S$-wave dominated
  charmonium state with possible some $D$-wave components.}
  \footnotetext[3]{The numbers in parentheses indicate the approximate masses
  of the particles in the unit of MeV.
  The dominant components of the particles ${\psi}(3770)$, ${\psi}(4040)$ and
  ${\psi}(4160)$ are usually considered as the $1^{3}D_{1}$, $3^{3}S_{1}$
  and $2^{3}D_{1}$ states, respectively \cite{pdg,npa915.125,ijtp56.1892}.
  Here, the spectroscopic notation $n^{2s+1}L_{J}$ is used, where $s$
  is the total spin of the quark-antiquark pair, and $J$ is the total
  angular momentum.}

  Phenomenologically, the nonleptonic $B$ meson weak decays have been
  studied carefully within the framework of the factorization hypothesis and
  the low-energy effective Hamiltonian \cite{rmp68.1125}.
  The naive factorization (NF) assumption \cite{plb73.418,npb133.315,zpc34.103}
  is usually employed in evaluating the nonleptonic $B$ meson decays,
  where the decay amplitudes in terms of hadronic matrix elements (HME) of
  the four-quark operators can be expressed as the product of two HME of
  the diquark currents based on Bjorken's color transparency argument
  \cite{npb11.325}. The diquark HME can be further parameterized
  by the decay constants or the hadron transition form factors.
  The NF hypothesis was verified experimentally to be successful for the
  class-I nonleptonic $B$ decays, but poor for the class-II ones.
  It is commonly believed that the characteristic space configuration of
  psions is compact, with the radius of $r$ ${\sim}$ $1/m_{c}$.
  The transverse separation between the two valence charm quarks should
  be very small.
  The massive psions from the $B$ meson decay could be regarded as color
  singlet states and factorized from the other system, although the velocity
  of psion might be not very large.
  The class-II $B$ ${\to}$ $J/{\psi}(1S)M$ decays have been studied based
  on the factorization assumption, such as in Refs. \cite{prd78.074030,
  prd68.034004,plb591.91,cpc34.937,prd89.094010,prd71.114008,epjc77.610,
  plb772.719,prd63.074011,prd65.094023}, where besides the factorizable
  contributions, the nonfactorizable contributions beyond the NF approximation
  are also taken into account to accommodate the discrepancies between
  the experimental data and the theoretical estimations.
  The $B$ ${\to}$ ${\psi}M$ decays provide a good place to check the
  factorization postulation and differentiate various
  theoretical treatments, such as the QCD factorization (QCDF) approach
  \cite{prl83.1914,npb591.313,plb488.46,npb606.245,plb509.263,
  prd64.014036,npb774.64,prd77.074013,1609.07430,npb675.333,npb795.1,
  npb822.172,npb832.109,jhep.2011.11.003,plb750.348,npb751.160,npb768.51,
  jhep.2007.05.019,npb794.154} based on the collinear approximation, and
  the perturbative QCD (pQCD) approach \cite{prl74.4388,plb348.597,prd52.3958,prd55.5577,
  prd56.1615,prd63.074006,plb504.6,prd63.054008,prd63.074009,epjc23.275}
  based on the collinear plus $k_{T}$ factorization supposition.

  It is well known that according to the quark model assignments, the spin-triplet
  charmonium states with different orbital angular momentum $L$ can have the same
  quantum numbers $J^{PC}$.
  The conservation of parity and angular momentum implies that the values of $L$
  for the mixed states can differ by two units at most.
  The psions near and above the open-charm threshold can
  be the admixtures of the $S$- and $D$-wave $c\bar{c}$ states
  \cite{zpc8.131,pan72.1206,prd79.094004,pan72.638,plb661.348,prd95.114031,
  prd96.014030,prd41.155,prd44.3562,prd64.094002,prd70.094001,pr41.1,zpc4.211}.
  The wave functions for the $S$-wave dominant state can receive the $D$-wave
  component and vice versa.
  Additionally, studies of the charmonium spectrum
  \cite{ijmpe22.1330026,prd95.034026,prd95.094021,epjc78.592}
  show that the mass of the $n^{3}S_{1}$ state is close to the mass
  of the $(n-1)^{3}D_{1}$ state.
  To the first-order approximation, the so-called $S$-$D$ wave mixing
  for psions refers mainly to the mixing between the $n^{3}S_{1}$
  and $(n-1)^{3}D_{1}$ charmonium states rather than the other
  states, and this has been used in the previous studies \cite{zpc8.131,pan72.1206,
  prd79.094004,pan72.638,plb661.348,prd95.114031,prd96.014030,prd41.155,
  prd44.3562,prd64.094002,prd70.094001,pr41.1,zpc4.211}.
  This $S$-$D$ wave mixing phenomenon might have certain effects on the
  production of psions in the $B$ ${\to}$ ${\psi}M$ decays.

  In this paper, we will investigate the $B_{u}$ ${\to}$ ${\psi}M$
  decays with the pQCD approach.
  Firstly, the electrically charged final meson $M$ should be easily identified
  by many specific detectors at the existing and future high energy
  colliders because of its track curve being saturated with the magnetic field.
  Secondly, the practicability of the pQCD approach can be checked with
  the class-II $B$ decays into final states containing the excited psions.
  Thirdly, the effects of the $S$-$D$ wave mixing among psions can be
  examined with the $B_{u}$ ${\to}$ ${\psi}M$ decays, without
  the disturbances from the mixing between the neutral $B$ mesons and
  without the pollution from the weak annihilation contributions.

  This paper is organized as follows.
  The theoretical framework and the amplitudes for the $B_{u}$ ${\to}$
  ${\psi}M$ decays are elaborated in Section \ref{sec02}.
  The numerical results and discussion are presented in Section \ref{sec03}.
  Finally, we give a short summary in Section \ref{sec04}.

  %%%%%%%%%%%%%%%%%%%%%%%%%%%%%%%%%%%%%%%%%%%%%%%%%%%%%%%%%%%
  \section{theoretical framework}
  \label{sec02}
  %%%%%%%%%%%%%%%%%%%%%%%%%%
  \subsection{The effective Hamiltonian}
  \label{sec0201}
  The $B_{u}$ ${\to}$
  ${\psi}M$ decays are actually induced by the weak interaction cascade processes
  $b$ ${\to}$ $c$ $+$ $W^{{\ast}-}$ ${\to}$ $c$ $+$ $\bar{c}\,q$ at the quark
  level within the standard model.
  Hence, some relevant energy scales are introduced theoretically, such as
  the infrared confinement scale ${\Lambda}_{\rm QCD}$ of the strong interactions,
  the mass $m_{b}$ for the decaying bottom quark, and the mass $m_{W}$
  for the virtual gauge boson $W^{{\ast}}$, with the clear size relation
  ${\Lambda}_{\rm QCD}$ ${\ll}$  $m_{b}$ ${\ll}$  $m_{W}$.
  The effective theory is usually used in practice to deal with the
  realistic multi-scale problems.
  With the operator product expansion and the renormalization group (RG) method,
  the effective Hamiltonian in charge of the $B_{u}$ ${\to}$ ${\psi}M$ decays
  can be written as \cite{rmp68.1125},
 %---------------------------------------------------------
   \begin{equation}
  {\cal H}_{\rm eff}\, =\, \frac{G_{F}}{\sqrt{2}}\,
   \sum\limits_{q=d,s}
   \Big\{ V_{cb}\,V_{cq}^{\ast} \sum\limits_{i=1}^{2}C_{i}({\mu})\,Q_{i}({\mu})
  - V_{tb}\,V_{tq}^{\ast} \sum\limits_{j=3}^{10}C_{j}({\mu})\,Q_{j}({\mu}) \Big\}
  +{\rm h.c.}
  \label{hamilton},
  \end{equation}
 %---------------------------------------------------------
  where the Fermi coupling constant $G_{F}$ ${\simeq}$
  $1.166{\times}10^{-5}\,{\rm GeV}^{-2}$ \cite{pdg}.
  $V_{pb}\,V_{pq}^{\ast}$ is the product of the Cabibbo-Kobayashi-Maskawa
  (CKM) matrix elements,
  satisfying the unitarity relation $V_{ub}\,V_{uq}^{\ast}$ $+$
  $V_{cb}\,V_{cq}^{\ast}$ $+$ $V_{tb}\,V_{tq}^{\ast}$ $=$ $0$.
  With the Wolfenstein parametrization, the CKM factors can be
  expanded as the power series of the parameter ${\lambda}$
  ${\approx}$ $0.2$\cite{pdg}. Up to ${\cal O}({\lambda}^{7})$,
  these CKM factors can be written as follows.
  %---------------------------------------------------------
  \begin{eqnarray}
  V_{cb}\,V_{cd}^{\ast}
  &=& -A\,{\lambda}^{3} +{\cal O}({\lambda}^{7})
  \label{vcbvcd}, \\
  %---------------------------------------------------------
  V_{tb}\,V_{td}^{\ast}
  &=& A\,{\lambda}^{3}\,\left(1-{\rho}+i\,{\eta}\right)
  +\frac{1}{2}\,A\,{\lambda}^{5}\,\left({\rho}-i\,{\eta}\right)
  +{\cal O}({\lambda}^{7})
  \label{vtbvtd}, \\
  %---------------------------------------------------------
  V_{cb}\,V_{cs}^{\ast}
  &=& A\,{\lambda}^{2}-\frac{1}{2}\,A\,{\lambda}^{4}
   - \frac{1}{8}\,A\,{\lambda}^{6}\,\left(1+4\,A^{2}\right)
  +{\cal O}({\lambda}^{7})
  \label{vcbvcs}, \\
  %---------------------------------------------------------
  V_{tb}\,V_{ts}^{\ast}
  &=& -V_{cb}\,V_{cs}^{\ast}-A\,{\lambda}^{4}\,\left({\rho}-i\,{\eta}\right)
  +{\cal O}({\lambda}^{7})
  \label{vtbvts}.
  \end{eqnarray}
  %---------------------------------------------------------
  The numerical values of the Wolfenstein parameters $A$, ${\lambda}$,
  ${\rho}$ and ${\eta}$ are listed in Table \ref{tab:input}.
  From the expression for $V_{pb}\,V_{pq}^{\ast}$ above, it is clearly
  seen that the weak phases for the $B_{u}$ ${\to}$ ${\psi}M$ decays
  are small, and thus result in a small direct $CP$ violation.

  The renormalization scale ${\mu}$ divides the physical contributions
  into the short- and long-distance parts.
  The physical contributions from the scale larger than ${\mu}$ are
  summarized in the Wilson coefficients $C_{i}$.
  The Wilson coefficients, $\vec{C}_{i}$ $=$ $\{C_{1},C_{2},{\cdots},C_{10}\}$,
  are calculable at the scale ${\mu}_{W}$ ${\sim}$ ${\cal O}(m_{W})$ with
  the perturbation theory, and then evolved to the characteristic scale
  ${\mu}_{b}$ ${\sim}$ ${\cal O}(m_{b})$ for the $b$ quark decay with
  the RG equation \cite{rmp68.1125},
  %-----------------------------------------------------
  \begin{equation}
  \vec{C}_{i}({\mu}_{b})\, =\, U({\mu}_{b},{\mu}_{W})\, \vec{C}_{i}({\mu}_{W})
  \label{Wilson},
  \end{equation}
  %-----------------------------------------------------
  where $U({\mu}_{b},{\mu}_{W})$ is the RG evolution matrix.
  The Wilson coefficients are independent of any process and
  have the same role as the universal gauge couplings.
  The expressions of the Wilson coefficients $\vec{C}_{i}(m_{W})$
  and $U({\mu}_{b},{\mu}_{W})$, including the next-to-leading order
  (NLO) corrections, can be found in Ref.\cite{rmp68.1125}.
  The physical contributions from the scale less than ${\mu}$ are
  incorporated into the HME,
  ${\langle}{\psi}M{\vert}Q_{i}{\vert}B_{u}{\rangle}$,
  where the local four-quark operators $Q_{i}$ are sandwiched
  between the initial and final hadron states.
  The operators are expressed as follows.
 %-----------------------------------------------------
  \begin{equation}
  Q_{1} \, =\, \bar{c}_{\alpha}\,{\gamma}_{\mu}\,(1-{\gamma}_{5})\,b_{\alpha}\
            \bar{q}_{\beta}\,{\gamma}^{\mu}\,(1-{\gamma}_{5})\,c_{\beta}
  \label{operator:q1},
  \end{equation}
 %-----------------------------------------------------
  \begin{equation}
  Q_{2} \, =\, \bar{c}_{\alpha}\,{\gamma}_{\mu}\,(1-{\gamma}_{5})\,b_{\beta}\
            \bar{q}_{\beta}\,{\gamma}^{\mu}\,(1-{\gamma}_{5})\,c_{\alpha}
  \label{operator:q2},
  \end{equation}
 %-----------------------------------------------------
  \begin{equation}
  Q_{3} \, =\, \sum\limits_{q^{\prime}}
            \bar{q}_{\alpha}\,{\gamma}_{\mu}\,(1-{\gamma}_{5})\,b_{\alpha}\
            \bar{q}^{\prime}_{\beta}\,{\gamma}^{\mu}\,(1-{\gamma}_{5})\,q^{\prime}_{\beta}
  \label{operator:q3},
  \end{equation}
 %-----------------------------------------------------
  \begin{equation}
  Q_{4} \, =\, \sum\limits_{q^{\prime}}
            \bar{q}_{\alpha}\,{\gamma}_{\mu}\,(1-{\gamma}_{5})\,b_{\beta}\
            \bar{q}^{\prime}_{\beta}\,{\gamma}^{\mu}\,(1-{\gamma}_{5})\,q^{\prime}_{\alpha}
  \label{operator:q4},
 \end{equation}
 %-----------------------------------------------------
  \begin{equation}
  Q_{5} \, =\, \sum\limits_{q^{\prime}}
            \bar{q}_{\alpha}\,{\gamma}_{\mu}\,(1-{\gamma}_{5})\,b_{\alpha}\
            \bar{q}^{\prime}_{\beta}\,{\gamma}^{\mu}\,(1+{\gamma}_{5})\,q^{\prime}_{\beta}
  \label{operator:q5},
  \end{equation}
 %-----------------------------------------------------
  \begin{equation}
  Q_{6} \, =\, \sum\limits_{q^{\prime}}
            \bar{q}_{\alpha}\,{\gamma}_{\mu}\,(1-{\gamma}_{5})\,b_{\beta}\
            \bar{q}^{\prime}_{\beta}\,{\gamma}^{\mu}\,(1+{\gamma}_{5})\,q^{\prime}_{\alpha}
  \label{operator:q6},
  \end{equation}
 %-----------------------------------------------------
  \begin{equation}
  Q_{7} \, =\, \sum\limits_{q^{\prime}}\frac{3}{2}e_{q^{\prime}}\,
            \bar{q}_{\alpha}\,{\gamma}_{\mu}\,(1-{\gamma}_{5})\,b_{\alpha}\
            \bar{q}^{\prime}_{\beta}\,{\gamma}^{\mu}\,(1+{\gamma}_{5})\,q^{\prime}_{\beta}
  \label{operator:q7},
  \end{equation}
 %-----------------------------------------------------
  \begin{equation}
  Q_{8} \, =\, \sum\limits_{q^{\prime}}\frac{3}{2}e_{q^{\prime}}\,
            \bar{q}_{\alpha}\,{\gamma}_{\mu}\,(1-{\gamma}_{5})\,b_{\beta}\
            \bar{q}^{\prime}_{\beta}\,{\gamma}^{\mu}\,(1+{\gamma}_{5})\,q^{\prime}_{\alpha}
  \label{operator:q8},
  \end{equation}
 %-----------------------------------------------------
  \begin{equation}
  Q_{9} \, =\, \sum\limits_{q^{\prime}}\frac{3}{2}e_{q^{\prime}}\,
            \bar{q}_{\alpha}\,{\gamma}_{\mu}\,(1-{\gamma}_{5})\,b_{\alpha}\
            \bar{q}^{\prime}_{\beta}\,{\gamma}^{\mu}\,(1-{\gamma}_{5})\,q^{\prime}_{\beta}
  \label{operator:q9},
  \end{equation}
 %-----------------------------------------------------
  \begin{equation}
  Q_{10} \, =\, \sum\limits_{q^{\prime}}\frac{3}{2}e_{q^{\prime}}\,
            \bar{q}_{\alpha}\,{\gamma}_{\mu}\,(1-{\gamma}_{5})\,b_{\beta}\
            \bar{q}^{\prime}_{\beta}\,{\gamma}^{\mu}\,(1-{\gamma}_{5})\,q^{\prime}_{\alpha}
  \label{operator:q10},
  \end{equation}
 %-----------------------------------------------------
  where $Q_{1,2}$ are the tree operators originating from the
  $W$-boson emission; $Q_{3,{\cdots},6}$ and $Q_{7,{\cdots},10}$ are
  the QCD and electroweak penguin operators, respectively;
  $(q_{1}\,q_{2})_{V{\pm}A}$ $=$ $q_{1}\,{\gamma}_{\mu}(1{\pm}{\gamma}_{5})\,q_{2}$;
  ${\alpha}$ and ${\beta}$ are color indices, i.e., the QCD corrections are
  considered; $q^{\prime}$ denotes all the active quarks at the scale of
  ${\cal O}(m_{b})$, i.e., $q^{\prime}$ $=$ $u$, $d$, $c$, $s$, $b$;
  and $e_{q^{\prime}}$ is the fractional electric charge of the quark
  $q^{\prime}$ in the unit of ${\vert}e{\vert}$.
  To obtain the decay amplitudes, the proper calculation of the HME
  ${\langle}{\psi}M{\vert}Q_{i}{\vert}B_{u}{\rangle}$
  will be the focus of the current research.

  %%%%%%%%%%%%%%%%%%%%%%%%%%
  \subsection{Hadronic matrix elements}
  \label{sec0202}
  The participation of the strong interaction greatly complicates the theoretical
  calculation of HME for the nonleptonic $B$ weak decays in a reliable way,
  because of the entanglement between the perturbative and nonperturbative
  contributions.
  To evaluate the nonfactorizable contributions to HME beyond the NF
  approximation \cite{plb73.418,npb133.315,zpc34.103}, many QCD-inspired
  phenomenological approaches, such as the QCDF \cite{prl83.1914,npb591.313,
  plb488.46,npb606.245,plb509.263,prd64.014036,npb774.64,prd77.074013,
  1609.07430,npb675.333,npb795.1,npb822.172,npb832.109,jhep.2011.11.003,
  plb750.348,npb751.160,npb768.51,jhep.2007.05.019,npb794.154}
  and pQCD \cite{prl74.4388,plb348.597,prd52.3958,prd55.5577,prd56.1615,
  prd63.074006,plb504.6,prd63.054008,prd63.074009,epjc23.275} approaches,
  have been developed recently based on the framework proposed by Lepage and
  Brodsky \cite{prd22.2157}.
  The short- and long-distance contributions are effectively coordinated,
  and the HME are written as the
  convolution of the universal wave functions (WFs) reflecting the
  nonperturbative contributions with the process-dependent hard scattering
  amplitudes containing perturbative contributions. With the pQCD approach,
  it is supposed that the final $M$ meson should be energetic in the rest
  frame of the initial $B_{u}$ meson. The soft spectator quark of
  the $B_{u}$ meson, carrying momentum of ${\cal O}({\Lambda}_{\rm QCQ})$, should
  be kicked by one hard gluon so that the spectator quark can move
  as fast as the light quark from the bottom quark weak decay and then
  be incorporated into the color-singlet $M$ meson. That means the
  spectator quark should interact with other quarks via one hard gluon
  exchange, as shown in Fig.\ref{fig:fey}.
  In the practical calculation, in order to circumvent the endpoint singularities
  appearing in the collinear approximation \cite{npb606.245,plb509.263,prd64.014036,
  npb774.64}, the pQCD approach suggests \cite{prl74.4388,plb348.597,prd52.3958}
  retaining the transverse momentum of the valence quarks and simultaneously
  introducing the Sudakov factors for all participant meson WFs to further
  depress the nonperturbative contributions.
  Finally, the pQCD decay amplitudes are divided into three parts \cite{plb348.597,
  prd52.3958,prd55.5577,prd56.1615,prd63.074006,plb504.6,prd63.054008,prd63.074009,epjc23.275}:
  the hard contributions enclosed by the Wilson coefficients $C_{i}$, the bottom quark
  scattering amplitudes ${\cal H}_{i}$, and the nonperturbative contributions absorbed into
  the mesonic WFs ${\Phi}_{i}$.
  The general form is a multidimensional integral,
  %-----------------------------------------------------
  \begin{equation}
 {\cal A}_{i}\ {\propto}\ {\int}\, {\prod_j}dx_{j}\,db_{j}\,
  C_{i}(t_{i})\,{\cal H}_{i}(t_{i},x_{j},b_{j})\,{\Phi}_{j}(x_{j},b_{j})\,e^{-S_{j}}
  \label{hadronic},
  \end{equation}
  %-----------------------------------------------------
  where $x_{j}$ is the longitudinal momentum fraction of the valence quarks;
  $b_{j}$ is the conjugate variable of the transverse momentum $k_{jT}$;
  $t_{i}$ is a typical scale; $e^{-S_{j}}$ is the Sudakov factor.
  In the numerical evaluations, besides the effective suppression on the
  long-distance contributions from the Sudakov factor, the scale $t_{i}$
  is usually chosen to be the maximum virtuality of all the internal
  particles, as shown in Eq.(\ref{scale-ti}), to further guarantee that
  the perturbative calculation of scattering amplitudes is practicable.

  %%%%%%%%%%%%%%%%%%%%%%%%%%
  \subsection{Kinematic variables}
  \label{sec0203}
  In the heavy quark limit, the light quark from the bottom quark
  decay is assumed to fly quickly away from the interaction point
  at near the speed of light.
  The light-cone dynamics can be used to describe the relativistic system.
  The relations between the four-dimensional space-time
  coordinates ($x^{0}$, $x^{1}$, $x^{2}$, $x^{3}$) $=$ ($t$, $x$, $y$, $z$)
  and the light-cone coordinates ($x^{+}$, $x^{-}$, $x_{\perp}$)
  are defined as $x^{\pm}$ $=$ $(x^{0}{\pm}x^{3})/\sqrt{2}$ and
  $x_{\perp}$ $=$ ($x^{1}$, $x^{2}$).
  The planes of $x^{\pm}$ $=$ $0$ are called the light-cone.
  The scalar product of any two vectors is given by
  $a{\cdot}b$ $=$ $a_{\mu}b^{\mu}$ $=$ $a^{+}b^{-}$ $+$ $a^{-}b^{+}$
  $-$ $a_{\perp}{\cdot}b_{\perp}$.
  In the rest frame of the $B_{u}$ meson, the final ${\psi}$
  and $M$ mesons move in the opposite direction.
  The light-cone kinematic variables are defined as follows.
  %------------------------------------
  \begin{equation}
  p_{B}\, =\, p_{1}\, =\, \frac{m_{1}}{\sqrt{2}}(1,1,0)
  \label{kine-p1},
  \end{equation}
  %------------------------------------
  \begin{equation}
  p_{\psi}\, =\, p_{2}\, =\, (p_{2}^{+},p_{2}^{-},0)
  \label{kine-p2},
  \end{equation}
  %------------------------------------
  \begin{equation}
  p_{M}\, =\, p_{3}\, =\, (p_{3}^{-},p_{3}^{+},0)
  \label{kine-p3},
  \end{equation}
  %------------------------------------
  \begin{equation}
  k_{i}\, =\, x_{i}\,p_{i}+(0,0,k_{iT})
  \label{kine-ki},
  \end{equation}
  %------------------------------------
  \begin{equation}
  p_{i}^{\pm}\, =\, (E_{i}\,{\pm}\,p_{\rm cm})/\sqrt{2}
  \label{kine-pipm},
  \end{equation}
  %------------------------------------
  \begin{equation}
  t \, =\, 2\,p_{1}{\cdot}p_{2}
  \, =\,2\,m_{1}\,E_{2}
  \label{kine-t},
  \end{equation}
  %------------------------------------
  \begin{equation}
  u \, =\, 2\,p_{1}{\cdot}p_{3}
  \, =\,2\,m_{1}\,E_{3}
  \label{kine-u},
  \end{equation}
  %------------------------------------
  \begin{equation}
  s \, =\, 2\,p_{2}{\cdot}p_{3}
  \label{kine-s},
  \end{equation}
  %------------------------------------
  \begin{equation}
  s\,t +s\,u-t\,u \ =\ 4\,m_{1}^{2}\,p_{\rm cm}^{2}
  \label{kine-pcm},
  \end{equation}
  %------------------------------------
  where the subscript $i$ $=$ $1$, $2$, $3$ on variables (including
  the mass $m_{i}$, momentum $p_{i}$ and energy $E_{i}$) correspond to
  the $B_{u}$, ${\psi}$ and $M$ mesons, respectively.
  The parameters $k_{i}$, $x_{i}$, $k_{iT}$ are the momentum, the longitudinal
  momentum fraction, and the transverse momentum of the valence antiquark, respectively.
  $p_{\rm cm}$ is the center-of-mass momentum of the final states.
  The notations of these momenta are displayed in Fig.\ref{fig:fey}(a).

  %%%%%%%%%%%%%%%%%%%%%%%%%%
  \subsection{Wave functions}
  \label{sec0204}
  The wave functions and/or distribution amplitudes (DAs) are the essential
  ingredient in the master pQCD formula of Eq.(\ref{hadronic}).
  Although nonperturbative, the WFs and DAs are generally considered to be
  universal for any process. The WFs and DAs determined by nonperturbative
  methods or extracted from data can be employed here to make predictions.
  Following the notations in Refs. \cite{npb529.323,prd65.014007,jhep9901.010,
  jhep0703.069,jhep0605.004,prd92.074028,plb751.171,plb752.322,prd96.036010},
  the WFs in question are defined as follows.
  %-----------------------------------------------------
  \begin{equation}
 {\langle}0{\vert}\bar{u}_{i}(z)b_{j}(0){\vert}B_{u}^{-}(p){\rangle}\,
 =\, \frac{i\,f_{B}}{4} {\int}d^{4}k\,e^{-ik{\cdot}z}
  \Big\{ \Big[ \!\!\not{p}\, {\Phi}_{B}^{a}(k)+
  m_{B}\, {\Phi}_{B}^{p}(k) \Big]\, {\gamma}_{5} \Big\}_{ji}
  \label{wf-bu01},
  \end{equation}
  %-----------------------------------------------------
  \begin{equation}
 {\langle}{\psi}(p,{\epsilon}^{\parallel}){\vert}
  \bar{c}_{i}(z)c_{j}(0){\vert}0{\rangle}\, =\,
  \frac{f_{\psi}}{4}{\int}d^{4}k\,e^{+ik{\cdot}z}\,
  \Big\{ \!\!\not{\epsilon}^{\parallel}\, \Big[
   m_{\psi}\,{\Phi}_{\psi}^{v}(k)
  +\!\not{p}\,{\Phi}_{\psi}^{t}(k)
   \Big] \Big\}_{ji}
  \label{wf-cc01},
  \end{equation}
  %-----------------------------------------------------
  \begin{equation}
 {\langle}{\psi}(p,{\epsilon}^{\perp}){\vert}
  \bar{c}_{i}(z)c_{j}(0){\vert}0{\rangle}\, =\,
  \frac{f_{\psi}}{4}{\int}d^{4}k\,e^{+ik{\cdot}z}\,
  \Big\{ \!\!\not{\epsilon}^{\perp}\, \Big[
   m_{\psi}\,{\Phi}_{\psi}^{V}(k)
  +\!\not{p}\,{\Phi}_{\psi}^{T}(k)
   \Big] \Big\}_{ji}
  \label{wf-cc02},
  \end{equation}
  %-----------------------------------------------------
  \begin{eqnarray}
 {\langle}P(p){\vert}\bar{q}_{i}(z)q^{\prime}_{j}(0){\vert}0{\rangle}\,
 &=& \frac{1}{4}{\int}d^{4}k\,e^{+ik{\cdot}z}\,
  \Big\{ {\gamma}_{5}\Big[ \!\!\not{p}\,{\Phi}_{P}^{a}(k) +
 {\mu}_{P}\,{\Phi}_{P}^{p}(k)
  \nonumber \\ & & \qquad \qquad +
 {\mu}_{P}\,(\not{n}_{+}\!\!\not{n}_{-}-1)\,{\Phi}_{P}^{t}(k)
  \Big] \Big\}_{ji}
  \label{wf-p},
  \end{eqnarray}
  %-----------------------------------------------------
  \begin{equation}
 {\langle}V(p,{\epsilon}^{\parallel}){\vert}\bar{q}_{i}(z)q^{\prime}_{j}(0){\vert}0{\rangle}\,
 =\, \frac{1}{4}{\int}d^{4}k\,e^{+ik{\cdot}z}\,
  \Big\{ \!\!\not{\epsilon}^{\parallel}\,m_{V}\,{\Phi}_{V}^{v}(k)
  +\!\!\not{\epsilon}^{\parallel}\!\!\not{p}\,{\Phi}_{V}^{t}(k)
  -m_{V}\,{\Phi}_{V}^{s}(k) \Big\}_{ji}
  \label{wf-v-el},
  \end{equation}
  %-----------------------------------------------------
  \begin{eqnarray}
 {\langle}V(p,{\epsilon}^{\perp}){\vert}\bar{q}_{i}(z)q^{\prime}_{j}(0){\vert}0{\rangle}\,
 &=& \frac{1}{4}{\int}d^{4}k\,e^{+ik{\cdot}z}\,
  \Big\{ \!\!\not{\epsilon}^{\perp}\,m_{V}\,{\Phi}_{V}^{V}(k)
  +\!\!\not{\epsilon}^{\perp}\!\!\not{p}\,{\Phi}_{V}^{T}(k)
  \nonumber \\ & & \qquad +
  \frac{i\,m_{V}}{p{\cdot}n_{+}}\,{\gamma}_{5}\,{\varepsilon}_{{\mu}{\nu}{\alpha}{\beta}}
  {\gamma}^{\mu}\,{\epsilon}^{{\perp}{\nu}}\,p^{\alpha}\,
  n_{+}^{\beta}\,{\Phi}_{V}^{A}(k) \Big\}_{ji}
  \label{wfv-et},
  \end{eqnarray}
  %-----------------------------------------------------
  where $f_{B}$ and $f_{\psi}$ are the decay constants of the
  $B_{u}$ and ${\psi}$ mesons, respectively.
  ${\epsilon}^{\parallel}$ (${\epsilon}^{\perp}$) is the longitudinal
  (transverse) polarization vector.
  $n_{+}$ $=$ $(1,0,0)$ and $n_{-}$ $=$ $(0,1,0)$ are the positive and
  negative null light-cone vectors satisfying the conditions of
  $n_{\pm}^{2}$ $=$ $0$ and $n_{+}{\cdot}n_{-}$ $=$ $1$.
  The chiral parameter ${\mu}_{P}$ is given by
  \cite{jhep9901.010},
  %-----------------------------------------------------
  \begin{equation}
 {\mu}_{P}\, =\, \frac{m_{\pi}^{2}}{m_{u}+m_{d}}
          \, =\, \frac{m_{K}^{2}}{m_{u,d}+m_{s}}
          \, {\approx}\, (1.6{\pm}0.2)\,\text{GeV}
  \label{up}.
  \end{equation}
  %-----------------------------------------------------

  According to the twist classification in Refs. \cite{npb529.323,
  prd65.014007,jhep9901.010,jhep0703.069,jhep0605.004}, the WFs
  of ${\Phi}_{B,P}^{a}$ and ${\Phi}_{{\psi},V}^{v,T}$ are twist-2,
  while the WFs of ${\Phi}_{B,P}^{p,t}$ and ${\Phi}_{{\psi},V}^{t,s,V,A}$
  are twist-3. The WFs for the $nS$ and $nD$ psion states are given
  in Appendix \ref{cc-wfs}.
  In general, these mesonic WFs are the functions of two variables,
  the longitudinal momentum fractions $x_{i}$ and the transverse momentum
  $k_{iT}$ of the valence quarks. It is unanimously assumed with both the
  QCDF and pQCD approaches that outside the soft regions, the contributions
  from the transverse momentum can be neglected and the collinear approximation
  should work well \cite{prl83.1914,npb591.313,plb488.46,npb606.245,plb509.263,
  prd64.014036,npb774.64,prd77.074013,prl74.4388,plb348.597,prd52.3958,prd55.5577,
  prd56.1615,prd63.074006,plb504.6,prd63.054008,prd63.074009,epjc23.275}.
  One can obtain the corresponding DAs by integrating out the transverse
  momentum from the WFs.
  Near the endpoint regions where $x_{i}$ ${\to}$ $0$ or $1$, the collinear
  factorization approximation should no longer be valid \cite{plb488.46,
  npb606.245,plb509.263,prd64.014036}.
  The pQCD approach \cite{prl74.4388,plb348.597,prd52.3958} suggests that
  the effects of the transverse momentum cannot be overlooked.
  In addition, the valence quarks have different momentum fractions and
  velocities near the endpoint. The hadrons cannot be regarded as
  color transparent. The Sudakov factors should be introduced for the
  participating WFs in order to suppress the soft and nonperturbative
  contributions from the small $x_{i}$ and the large $k_{iT}$ regions
  \cite{prl74.4388,plb348.597,prd52.3958,prd55.5577,
  prd56.1615,prd63.074006,plb504.6,prd63.054008,prd63.074009,epjc23.275}.

  In our calculation, the expressions for the DAs involved are listed as follows
  \cite{prd65.014007,jhep9901.010,jhep0703.069,jhep0605.004,
  prd92.074028,plb751.171,plb752.322,prd96.036010}:
  %-----------------------------------------------------
   \begin{equation}
  {\phi}_{B}^{p}(x) \, =\, A\, {\exp}\Big\{
   -\frac{1}{8\,{\omega}_{1}^{2}}\,\Big(
    \frac{m_{u}^{2}}{x}
   +\frac{m_{b}^{2}}{\bar{x}} \Big) \Big\}
   \label{wave-bup},
   \end{equation}
  %-----------------------------------------------------
   \begin{equation}
  {\phi}_{B}^{a}(x)\, =\, B\, {\phi}_{B}^{p}(x)\, x\,\bar{x}
   \label{wave-bua},
   \end{equation}
  %-----------------------------------------------------
   \begin{equation}
  {\phi}_{{\psi}(1S)}^{v}(x) \, =\, C\, x\,\bar{x}\,
  {\exp}\Big\{ -\frac{1}{8\,{\omega}_{2}^{2}} \Big(
   \frac{m_{c}^{2}}{x}+\frac{m_{c}^{2}}{\bar{x}} \Big) \Big\}
   \label{wave-1s-v},
   \end{equation}
  %-----------------------------------------------------
   \begin{equation}
  {\phi}_{{\psi}(2S)}^{v}(x) \, =\, D\,{\phi}_{{\psi}(1S)}^{v}(x)
   \Big\{ 1+\frac{m_{c}^{2}}{ 2\,{\omega}_{2}^{2}\,x\,\bar{x} } \Big\}
   \label{wave-2s-v},
   \end{equation}
  %-----------------------------------------------------
   \begin{equation}
  {\phi}_{{\psi}(3S)}^{v}(x) \, =\, E\,{\phi}_{{\psi}(1S)}^{v}(x)
   \Big\{ \Big(1-\frac{m_{c}^{2}}{ 2\,{\omega}_{2}^{2}\,x\,\bar{x} }\Big)^{2}+6 \Big\}
   \label{wave-3s-v},
   \end{equation}
  %-----------------------------------------------------
   \begin{equation}
  {\phi}_{{\psi}(1D)}^{v}(x) \, =\, F\,{\phi}_{{\psi}(1S)}^{v}(x)
   \Big\{ 1+\frac{m_{c}^{2}}{ 8\,{\omega}_{2}^{2}\,x\,\bar{x} } \Big\}
   \label{wave-1d-v},
   \end{equation}
  %-----------------------------------------------------
   \begin{equation}
  {\phi}_{{\psi}(2D)}^{v}(x) \, =\, G\,{\phi}_{{\psi}(1S)}^{v}(x)
   \Big\{ \Big(1+\frac{m_{c}^{2}}{ {\omega}_{2}^{2}\,x\,\bar{x} }\Big)^{2}+15 \Big\}
   \label{wave-2d-v},
   \end{equation}
  %-----------------------------------------------------
   \begin{equation}
  {\phi}_{{\psi}}^{t}(x) \, =\, H\,{\phi}_{{\psi}}^{v}(x)\,{\xi}^{2}/(x\,\bar{x})
   \label{wave-v-t},
   \end{equation}
  %-----------------------------------------------------
   \begin{equation}
  {\phi}_{{\psi}}^{V}(x) \, =\, I\,{\phi}_{{\psi}}^{v}(x)\,(1+{\xi}^{2})/(x\,\bar{x})
   \label{wave-v-V},
   \end{equation}
  %-----------------------------------------------------
   \begin{equation}
  {\phi}_{{\psi}}^{T}(x) \, =\, J\,{\phi}_{{\psi}}^{v}(x)
   \label{wave-v-T},
   \end{equation}
  %-----------------------------------------------------
   \begin{equation}
  {\phi}_{P}^{a}(x)\, =\, i\,f_{P}\,6\,x\,\bar{x}\,
   \sum\limits_{i=0}^{n} a^{P}_{i}\,C_{i}^{3/2}({\xi})
   \label{da-pa},
   \end{equation}
  %-----------------------------------------------------
   \begin{equation}
 {\phi}_{V}^{v}(x) \, =\, f_{V}\,6\,x\,\bar{x}\,
  \sum\limits_{i=0}^{n} a^{\parallel}_{i}\,C_{i}^{3/2}({\xi})
  \label{da-rho-v},
  \end{equation}
  %-----------------------------------------------------
  \begin{equation}
 {\phi}_{V}^{T}(x) \, =\, f_{V}^{T}\,6\,x\,\bar{x}\,
  \sum\limits_{i=0}^{n} a^{\perp}_{i}\,C_{i}^{3/2}({\xi})
  \label{da-rho-T},
  \end{equation}
  %-----------------------------------------------------
  \begin{equation}
 {\phi}_{P}^{p}(x) \, =\, +i\,f_{P}\, C_{0}^{1/2}({\xi})
  \label{da-pp},
  \end{equation}
  %-----------------------------------------------------
  \begin{equation}
 {\phi}_{P}^{t}(x) \, =\, -i\,f_{P}\, C_{1}^{1/2}({\xi})
  \label{da-pt},
  \end{equation}
  %-----------------------------------------------------
  \begin{equation}
 {\phi}_{V}^{t}(x) \, =\, +3\, f_{V}^{T}\,{\xi}^{2}
  \label{da-rho-t},
  \end{equation}
  %-----------------------------------------------------
  \begin{equation}
 {\phi}_{V}^{s}(x) \, =\, -3\, f_{V}^{T}\,{\xi}
  \label{da-rho-s},
  \end{equation}
  %-----------------------------------------------------
  \begin{equation}
 {\phi}_{V}^{V}(x) \, =\, +\frac{3}{4}\,f_{V}\,(1+{\xi}^{2})
  \label{da-rho-V},
  \end{equation}
  %-----------------------------------------------------
  \begin{equation}
 {\phi}_{V}^{A}(x) \, =\, -\frac{3}{2}\,f_{V}\,{\xi}
  \label{da-rho-A}.
  \end{equation}
  %-----------------------------------------------------
  where $x$ and $\bar{x}$ $=$ $1$ $-$ $x$ are the momentum fractions of the
  valence quarks. The variable ${\xi}$ $=$ $x$ $-$ $\bar{x}$.
  The parameter ${\omega}_{i}$ determines the average transverse momentum
  of partons and ${\omega}_{i}$ ${\simeq}$ $m_{i}\,{\alpha}_{s}$
  \cite{prd92.074028,plb751.171,plb752.322,prd96.036010,prd46.4052,prd51.1125,rmp77.1423}.
  The parameters $A$, ${\cdots}$, $J$ in Eqs.(\ref{wave-bup}-\ref{wave-v-T})
  are the normalization coefficients.
  The DAs of Eqs.(\ref{wave-bup}-\ref{wave-v-T}) satisfy the normalization conditions,
  %-----------------------------------------------------
   \begin{equation}
  {\int}_{0}^{1}dx\,{\phi}_{B}^{a,p}(x) =1
   \label{wave-b-n},
   \end{equation}
  %-----------------------------------------------------
   \begin{equation}
  {\int}_{0}^{1}dx\,{\phi}_{{\psi}(nL)}^{v,t,V,T}(x)=1
   \label{wave-psi-n}.
   \end{equation}
  %-----------------------------------------------------

  The DAs of Eqs.(\ref{da-pa}-\ref{da-rho-A}) are the normalized expressions.
  The parameter $f_{P}$ is the decay constant for the pseudoscalar meson $P$.
  The parameters $f_{V}$ and $f_{V}^{T}$ are the longitudinal and transverse
  decay constants for the vector meson $V$.
  The nonperturbative parameters $a_{i}^{P,{\parallel},{\perp}}$
  are the Gegenbauer moments, and with $a_{0}^{P,{\parallel},{\perp}}$
  $=$ $1$ for the asymptotic forms.
  The Gegenbauer polynomials $C_{i}^{j}({\xi})$ are expressed as follows:
  %-----------------------------------------------------
   \begin{eqnarray}
   & & C_{0}^{j}({\xi})\, =\, 1
   \label{eq-c0}, \\
   & & C_{1}^{j}({\xi})\, =\, 2\,j\,{\xi}
   \label{eq-c1}, \\
   & & C_{2}^{j}({\xi})\, =\, 2\,j\,(j+1)\,{\xi}^{2}-j
   \label{eq-c2}, \\
   & & ...... \nonumber
   \end{eqnarray}
  %-----------------------------------------------------
  %-----------------------------------------------------
  \begin{figure}[h]
  \includegraphics[width=0.90\textwidth,bb=75 400 530 710]{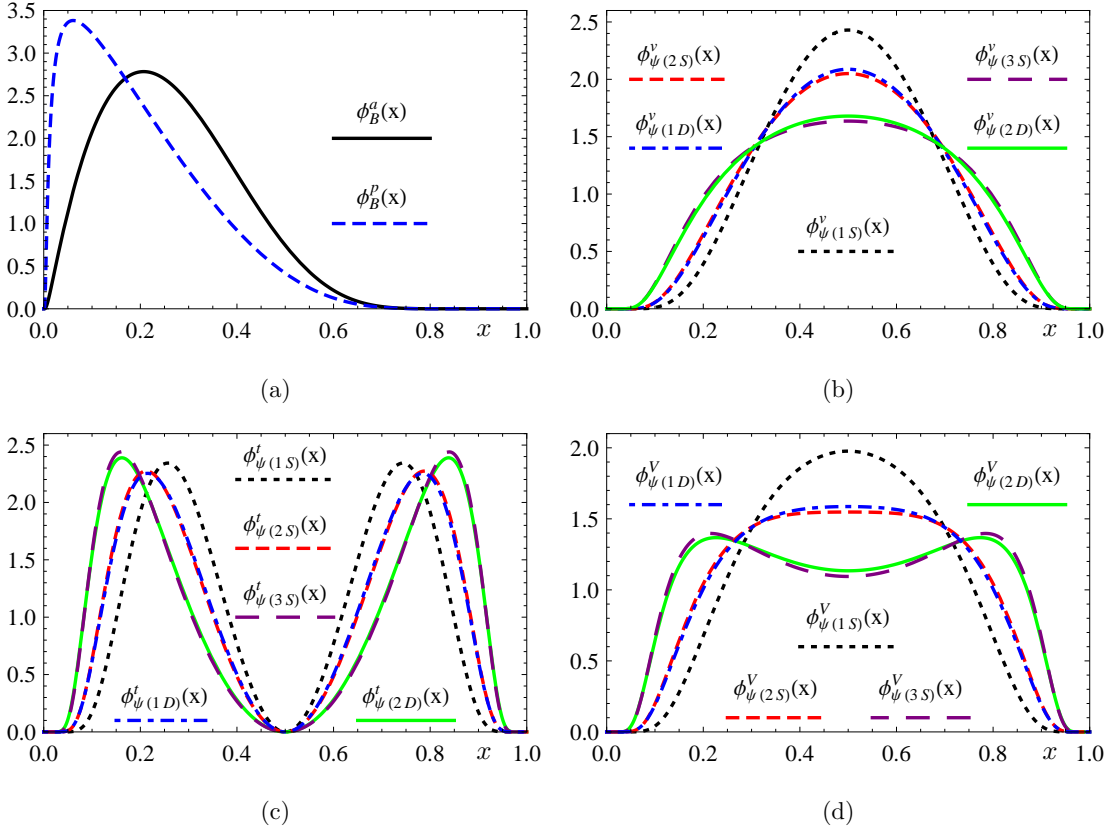}
  \caption{The normalized DAs of ${\phi}_{B}^{a,p}(x)$ and
  ${\phi}_{\psi}^{v,t,V}(x)$ (vertical axis) versus the parton
  momentum fraction $x$ (horizontal axis).}
  \label{fig:da}
  \end{figure}
  %-----------------------------------------------------

  A distinguishing feature of the DAs in Eqs.(\ref{wave-bup}-\ref{wave-v-T})
  is the exponential functions.
  These exponential factors are proportional to the ratio of $m_{i}^{2}/x_{i}$,
  so that the shape lines of DAs in Eqs.(\ref{wave-bup}-\ref{wave-v-T}) are generally
  consistent with the seemingly plausible suspicion that the momentum fractions $x_{i}$
  are shared by the valence quarks according to the quark mass $m_{i}$.
  In addition, the DAs will approach zero when $x_{i}$ ${\to}$ $0$ and $1$, due to the
  effective cutoff of the endpoint contributions from the exponential functions.
  The curves of the normalized DAs ${\phi}_{B}^{a,p}(x)$ and ${\phi}_{\psi}^{v,t,V}(x)$
  in Eqs.(\ref{wave-bup}-\ref{wave-v-T}) versus the parton momentum fraction $x$
  are shown in Fig.\ref{fig:da}.
  It is seen that
  (1) the parton momentum fraction of the spectator quark in the $B_{u}$ meson
  peaks in the $x$ $<$ $0.4$ region;
  (2) the DAs of ${\phi}_{\psi}^{v,t,V}(x)$ are symmetric with respect to
  the $x$ ${\leftrightarrow}$ $\bar{x}$ transformation; and
  (3) the difference between the DAs for the $2S$ and $1D$ psion states
  (and the $3S$ and $2D$ psion states) is subtle.

  %-----------------------------------------------------
   \begin{table}[ht]
   \caption{Some properties of the psion resonances \cite{pdg},
   where ${\Gamma}$ denotes the full decay width; ${\cal B}r_{ee}$
   and ${\Gamma}_{ee}$ denote the branching ratio and partial
   width for the pure leptonic ${\psi}$ ${\to}$ $e^{+}e^{-}$
   decay; $f_{\psi}$ is the decay constant obtained with Eq.(\ref{eq:fpsi2});
   ${\alpha}_{s}(m_{\psi})$ is the QCD coupling at the
   scale ${\mu}$ $=$ $m_{\psi}$.}
   \label{tab:psi}
   \begin{ruledtabular}
   \begin{tabular}{ccccccc}
    meson
  & mass (MeV)
  & ${\Gamma}$ (keV)
  & ${\cal B}r_{ee}$
  & ${\Gamma}_{ee}$ (keV)
  & $f_{\psi}$ (MeV)
  & ${\alpha}_{s}(m_{\psi})$ \\ \hline
    ${\psi}(2S)$
  & $3686.097{\pm}0.025$
  & $296{\pm}8$
  & $(7.89{\pm}0.17){\times}10^{-3}$
  & $2.34{\pm}0.04$
  & $358.8{\pm}3.1$
  & $0.227$ \\
    ${\psi}(3770)$
  & $3773.13{\pm}0.35$
  & $(27.2{\pm}1.0){\times}10^{3}$
  & $(9.6{\pm}0.7){\times}10^{-6}$
  & $0.262{\pm}0.018$
  & $121.2{\pm}4.2$
  & $0.225$ \\
    ${\psi}(4040)$
  & $4039{\pm}1$
  & $(80{\pm}10){\times}10^{3}$
  & $(1.07{\pm}0.16){\times}10^{-5}$
  & $0.86{\pm}0.07$
  & $225.4{\pm}9.4$
  & $0.220$ \\
    ${\psi}(4160)$
  & $4191{\pm}5$
  & $(70{\pm}10){\times}10^{3}$
  & $(6.9{\pm}3.3){\times}10^{-6}$
  & $0.48{\pm}0.22$
  & $170.9{\pm}45.2$
  & $0.217$
   \end{tabular}
   \end{ruledtabular}
   \end{table}
  %-----------------------------------------------------

  Some properties of the psion resonances are collected in Table \ref{tab:psi},
  where the decay constant $f_{\psi}$ is defined by
  ${\langle}0{\vert}\bar{c}\,{\gamma}^{\mu}\,c{\vert}{\psi}{\rangle}$ $=$
  $f_{\psi}\,m_{\psi}\,{\epsilon}_{\psi}^{\mu}$ and can be extracted from the
  electronic ${\psi}$ ${\to}$ $e^{+}e^{-}$ decay through the formula including
  the QCD radiative corrections \cite{plb57.455,npb105.125,prd19.1517,prd20.1175,
  prd21.203,zpc5.239,zpc8.131,pan72.1206,prd79.094004,pan72.638},
  %-----------------------------------------------------
   \begin{equation}
  {\Gamma}_{ee}\, =\,
  {\Gamma}({\psi}{\to}e^{+}e^{-})\, =\,
   \frac{16\,{\pi}}{27}\,{\alpha}_{\rm QED}^{2}(m_{\psi})\,
   \frac{f_{\psi}^{2}}{m_{\psi}}\,\Big\{ 1
  -\frac{16}{3\,{\pi}}\,{\alpha}_{s}(m_{\psi}) \Big\}
   \label{eq:fpsi2},
   \end{equation}
  %-----------------------------------------------------
  where the RG evolution equation for the coupling ${\alpha}_{\rm QED}$
  (${\alpha}_{s}$) of the electromagnetic (strong) interactions is given
  in Ref.\cite{prd59.054008} (Ref.\cite{rmp68.1125}).
  In our calculation, the one-loop leptonic
  contributions to ${\alpha}_{\rm QED}$ are considered with the initial
  value ${\alpha}_{\rm QED}(m_{W})$ $=$ $1/128$ resulting in
  ${\alpha}_{\rm QED}(m_{\psi})$ ${\sim}$ $1/131$, and the NLO contributions
  to the coupling ${\alpha}_{s}$ of the strong interactions are considered
  with the initial value ${\alpha}_{s}(m_{Z})$ $=$ $0.1182$ \cite{pdg}.
  It is seen from Table \ref{tab:psi} that there exist differences in the
  dielectric psion decay widths, which is assumed to be accommodated appropriately
  with the interferences between the $S$- and $D$- states \cite{zpc8.131,pan72.1206,
  prd79.094004,pan72.638,plb661.348,prd95.114031,prd96.014030,prd41.155,
  prd44.3562,prd64.094002,prd70.094001,pr41.1,zpc4.211}.
  Although with nearly the same shape lines for the $2S$- and $1D$-wave
  (and the $3S$- and $2D$-wave) psion DAs (see Fig.\ref{fig:da}),
  the differences in the decay constants
  might have an influence on the $B_{u}$ ${\to}$ ${\psi}M$ decays due to the
  $S$-$D$ mixing.
  In this paper, the $S$-$D$ wave mixing effects on the $B_{u}$ ${\to}$ ${\psi}M$
  decays are investigated.
  The physical psion mesons are the admixtures of the $S$- and $D$- states
  \cite{zpc8.131,pan72.1206,
  prd79.094004,pan72.638,plb661.348,prd95.114031,prd96.014030,prd41.155,
  prd44.3562,prd64.094002,prd70.094001,pr41.1,zpc4.211},
  %-----------------------------------------------------
   \begin{equation}
   \bigg( \begin{array}{c} {\psi}(3686) \\ {\psi}(3770) \end{array} \bigg)
   \, =\, \bigg( \begin{array}{cc} {\cos}{\theta}_{1} & {\sin}{\theta}_{1} \\
 -{\sin}{\theta}_{1} & {\cos}{\theta}_{1} \end{array} \bigg)\,
   \bigg( \begin{array}{c} {\psi}(2S) \\ {\psi}(1D) \end{array} \bigg)
   \label{eq:mix01},
   \end{equation}
  %-----------------------------------------------------
   \begin{equation}
   \bigg( \begin{array}{c} {\psi}(4040) \\ {\psi}(4160) \end{array} \bigg)
   \, =\, \bigg( \begin{array}{cc} {\cos}{\theta}_{2} & {\sin}{\theta}_{2} \\
 -{\sin}{\theta}_{2} & {\cos}{\theta}_{2} \end{array} \bigg)\,
   \bigg( \begin{array}{c} {\psi}(3S) \\ {\psi}(2D) \end{array} \bigg)
   \label{eq:mix02},
   \end{equation}
  %-----------------------------------------------------
  where the subscript $i$ of the $S$-$D$ mixing angle ${\theta}_{i}$ corresponds
  to the radial quantum number $n$ of the ${\psi}(nD)$ states.
  There are two sets of possible ranges for the value of the $2S$-$1D$ mixing
  angle \cite{pan72.1206,prd79.094004,pan72.638,plb661.348,
  prd95.114031,prd96.014030,prd41.155,prd44.3562,prd64.094002,prd70.094001}\footnotemark[4], i.e.,
  ${\theta}_{1}$ ${\approx}$ $-10^{\circ}$ ${\sim}$ $-14^{\circ}$ and
  ${\theta}_{1}$ ${\approx}$ $+25^{\circ}$ ${\sim}$ $+30^{\circ}$.
  The possible value of the $3S$-$2D$ mixing angle is
  ${\theta}_{2}$ ${\approx}$ $-35^{\circ}$
  \cite{prd79.094004,pan72.638,plb661.348,prd95.114031,prd96.014030}.
  As an approximation in the numerical computation,
  the values of ${\theta}_{1}$ ${\approx}$ $-(12{\pm}2)^{\circ}$ and
  $+(27{\pm}2)^{\circ}$ \cite{prd64.094002} and
  ${\theta}_{2}$ ${\approx}$ $-35^{\circ}$ \cite{pan72.638,plb661.348} will be used.
  The assumed mass relations are $m_{{\psi}(2S)}$ ${\approx}$ $m_{{\psi}(3686)}$,
  $m_{{\psi}(1D)}$ ${\approx}$ $m_{{\psi}(3770)}$, $m_{{\psi}(3S)}$ ${\approx}$ $m_{{\psi}(4040)}$
  and $m_{{\psi}(2D)}$ ${\approx}$ $m_{{\psi}(4160)}$.

  \footnotetext[4]{The possible values of the $2S$-$1D$ mixing angle are:
  ${\theta}_{1}$ ${\approx}$ $-10^{\circ}$ and $+30^{\circ}$ in Ref.\cite{prd41.155},
  ${\theta}_{1}$ ${\approx}$ $-13^{\circ}$ and $+26^{\circ}$ in Ref.\cite{prd44.3562},
  ${\theta}_{1}$ ${\approx}$ $-(12{\pm}2)^{\circ}$ and $+(27{\pm}2)^{\circ}$ in Ref.\cite{prd64.094002},
  ${\theta}_{1}$ ${\approx}$ $-12^{\circ}$ and $+25^{\circ}$ in Ref.\cite{prd79.094004},
  ${\theta}_{1}$ ${\approx}$ $-11^{\circ}$ in Ref.\cite{pan72.1206}.}

  %%%%%%%%%%%%%%%%%%%%%%%%%%
  \subsection{Decay amplitudes}
  \label{sec0205}
  %-----------------------------------------------------
  \begin{figure}[h]
  \includegraphics[width=0.90\textwidth,bb=90 625 515 710]{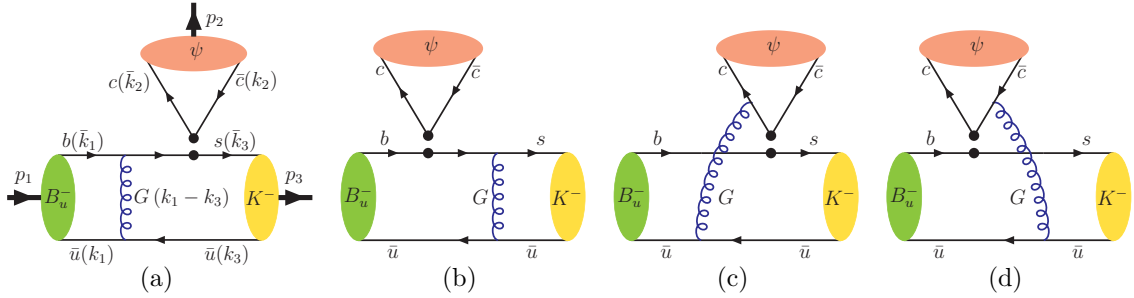}
  \caption{The Feynman diagrams for the $B_{u}^{-}$ ${\to}$ ${\psi}K^{-}$
   decay with the pQCD approach, where (a,b) and (c,d) are
   factorizable and nonfactorizable topologies, respectively.}
  \label{fig:fey}
  \end{figure}
  %-----------------------------------------------------

  Within the pQCD framework, the Feynman diagrams for the $B_{u}$ ${\to}$
  ${\psi}K$ decay are shown in Fig.\ref{fig:fey}. The spectator quark
  always interacts with one hard gluon in each subdiagram.
  The diagrams Fig.\ref{fig:fey}(a,b) are the factorizable emission
  topologies, where the gluons are exchanged between the initial $B_{u}$
  meson and the recoil $K$ meson.
  It is possible to completely isolate the emission psion particle from
  the $B_{u}K$ system, and hence the integral of the psion WFs will reduce
  to the psion decay constant.
  The diagrams Fig.\ref{fig:fey}(c,d) are the nonfactorizable emission
  topologies, where the gluons are exchanged between the psion particle and
  the $B_{u}K$ system, and hence no meson can escape from the
  interferences of other mesons.
  The diagrams Fig.\ref{fig:fey}(c,d) are also called the spectator scattering
  topologies with the QCDF approach \cite{npb591.313,plb488.46,npb606.245,
  plb509.263,prd64.014036,npb774.64}.
  The nonfactorizable HME can be written as the convolution integral of
  all the participating meson WFs.
  Compared with the factorizable contributions from Fig.\ref{fig:fey}(a,b),
  the nonfactorizable contributions from Fig.\ref{fig:fey}(c,d) are
  color-suppressed,
  which is quite similar to the cases between the external and
  internal $W$ emission topologies.

  After a direct calculation, the amplitudes for the $B_{u}$ ${\to}$ ${\psi}M$
  decays are written as follows:
  %-----------------------------------------------------
   \begin{eqnarray}
   \lefteqn{ {\cal A}(B_{u}{\to}{\psi}P)\, =\,
   \frac{{\pi}\,G_{F}\,C_{F}}{\sqrt{2}\,N_{c}}\, f_{B}\,f_{\psi}\,
   \Big\{ \big( V_{cb}V_{cd}^{\ast}\,{\delta}_{P,{\pi}}
  +V_{cb}V_{cs}^{\ast}\,{\delta}_{P,K} \big)\,\Big[
   a_{2}\,\big( {\cal A}_{a,P}^{LL}+{\cal A}_{b,P}^{LL} \big) }
   \nonumber \\ &+& C_{1}\,
   \big( {\cal A}_{c,P}^{LL}+{\cal A}_{d,P}^{LL} \big) \Big]
  -\big( V_{tb}V_{td}^{\ast}\,{\delta}_{P,{\pi}}
  +V_{tb}V_{ts}^{\ast}\,{\delta}_{P,K} \big)\, \Big[
   \big( a_{3}+a_{9} \big)\,
   \big( {\cal A}_{a,P}^{LL}+{\cal A}_{b,P}^{LL} \big)
   \nonumber \\ &+& \big( a_{5}+a_{7} \big)\,
   \big( {\cal A}_{a,P}^{LR}+{\cal A}_{b,P}^{LR} \big)
  +\big( C_{4}+C_{10} \big)\,
   \big( {\cal A}_{c,P}^{LL}+{\cal A}_{d,P}^{LL} \big)
   \nonumber \\ &+& \big( C_{6}+C_{8} \big)\,
   \big( {\cal A}_{c,P}^{LR}+{\cal A}_{d,P}^{LR} \big)
   \Big] \Big\}
   \label{eq:psi-p},
   \end{eqnarray}
  %-----------------------------------------------------
   \begin{equation}
  {\cal A}(B_{u}{\to}{\psi}V)\, =\,
  {\cal A}_{L}({\epsilon}^{\parallel}_{\psi}{\cdot}{\epsilon}^{\parallel}_{V})
 +{\cal A}_{N}\,({\epsilon}^{\perp}_{\psi}{\cdot}{\epsilon}^{\perp}_{V})
 +i\,{\cal A}_{T}\, {\varepsilon}_{{\mu}{\nu}{\alpha}{\beta}}\,
  {\epsilon}^{\mu}_{\psi}\, {\epsilon}^{\nu}_{V}\,p_{\psi}^{\alpha}\,p_{V}^{\beta}
   \label{eq:psi-v-01},
   \end{equation}
  %-----------------------------------------------------
   \begin{eqnarray}
   \lefteqn{ {\cal A}_{i}(B_{u}{\to}{\psi}V)\, =\,
   \frac{{\pi}\,G_{F}\,C_{F}}{\sqrt{2}\,N_{c}}\, f_{B}\,f_{\psi}\,
   \Big\{ \big( V_{cb}V_{cd}^{\ast}\,{\delta}_{V,{\rho}}
  +V_{cb}V_{cs}^{\ast}\,{\delta}_{V,K^{\ast}} \big)\,\Big[
   a_{2}\,\big( {\cal A}_{a,i}^{LL}+{\cal A}_{b,i}^{LL} \big) }
   \nonumber \\ &+& C_{1}\,
   \big( {\cal A}_{c,i}^{LL}+{\cal A}_{d,i}^{LL} \big) \Big]
  -\big( V_{tb}V_{td}^{\ast}\,{\delta}_{V,{\rho}}
  +V_{tb}V_{ts}^{\ast}\,{\delta}_{V,K^{\ast}} \big)\, \Big[
   \big( a_{3}+a_{9} \big)\,
   \big( {\cal A}_{a,i}^{LL}+{\cal A}_{b,i}^{LL} \big)
   \nonumber \\ &+& \big( a_{5}+a_{7} \big)\,
   \big( {\cal A}_{a,i}^{LR}+{\cal A}_{b,i}^{LR} \big)
  +\big( C_{4}+C_{10} \big)\,
   \big( {\cal A}_{c,i}^{LL}+{\cal A}_{d,i}^{LL} \big)
   \nonumber \\ &+& \big( C_{6}+C_{8} \big)\,
   \big( {\cal A}_{c,i}^{LR}+{\cal A}_{d,i}^{LR} \big)
   \Big] \Big\}, \qquad \quad \text{for}\ i\,=\,L,N,T
   \label{eq:psi-v-02}
   \end{eqnarray}
  %-----------------------------------------------------
   \begin{equation}
   a_{i}\, =\, \bigg\{ \begin{array}{lll}
   C_{i}+C_{i+1}/N_{c}, & \quad & \text{for odd}\ i \\
   C_{i}+C_{i-1}/N_{c}, & \quad & \text{for even}\ i
   \end{array}
   \label{eq:ai}
   \end{equation}
  %-----------------------------------------------------
  where the color factor $C_{F}$ $=$ $4/3$ and the color number $N_{c}$ $=$ $3$.
  For the amplitude building block ${\cal A}_{i,j}^{k}$, the subscript $i$
  corresponds to the subdiagram indices of Fig.\ref{fig:fey}; the subscript
  $j$ $=$ $P$, $L$, $N$, $T$ denotes the invariant polarization amplitudes, and
  the superscript $k$ refers to the two possible Dirac structures
  ${\Gamma}_{1}{\otimes}{\Gamma}_{2}$ of the operators
  $(\bar{q}_{1}q_{2})_{{\Gamma}_{1}}(\bar{q}_{3}q_{4})_{{\Gamma}_{2}}$,
  namely $k$ $=$ $LL$ for $(V-A){\otimes}(V-A)$ and $k$ $=$ $LR$
  for $(V-A){\otimes}(V+A)$.
  The explicit expressions of the building blocks ${\cal A}_{i,j}^{k}$
  are collected in Appendix \ref{block}.

  In addition, the amplitudes for the $B_{u}$ ${\to}$ ${\psi}V$ decays
  are conventionally expressed as the helicity amplitudes.
  The relation between the helicity amplitudes $H_{0,{\parallel},{\perp}}$
  and the scalar amplitudes ${\cal A}_{L,N,T}$ is
  \cite{prd66.054013,ijmpa31.1650146,npb911.890,prd95.036024}:
  %-----------------------------------------------------
   \begin{equation}
   H_{0}\ =\ {\cal A}_{L}({\epsilon}_{\psi}^{\parallel}{\cdot}{\epsilon}_{V}^{\parallel})
   \label{eq:psi-v-11},
   \end{equation}
  %-----------------------------------------------------
   \begin{equation}
   H_{\parallel}\ =\ \sqrt{2}\,{\cal A}_{N}
   \label{eq:psi-v-12},
   \end{equation}
  %-----------------------------------------------------
   \begin{equation}
   H_{\perp}\ =\ \sqrt{2}\,m_{B_{u}}\,p_{\rm cm}\, {\cal A}_{T}
   \label{eq:psi-v-13}.
   \end{equation}
  %-----------------------------------------------------

  %%%%%%%%%%%%%%%%%%%%%%%%%%%%%%%%%%
  \section{Numerical results and discussion}
  \label{sec03}
  In the rest frame of the $B_{u}$ meson, the branching
  ratios are defined as:
  %-----------------------------------------------------
   \begin{equation}
  {\cal B}r(B_{u}{\to}{\psi}P)\, =\,
   \frac{{\tau}_{B_{u}}}{8{\pi}}\,
   \frac{p_{\rm cm}}{m_{B_{u}}^{2}}\,
  {\vert}{\cal A}(B_{u}{\to}{\psi}P){\vert}^{2}
   \label{br-p},
   \end{equation}
  %-----------------------------------------------------
   \begin{equation}
  {\cal B}r(B_{u}{\to}{\psi}V)\, =\,
   \frac{{\tau}_{B_{u}}}{8{\pi}}\,
   \frac{p_{\rm cm}}{m_{B_{u}}^{2}}\, \Big\{
  {\vert}H_{0}{\vert}^{2}
 +{\vert}H_{\parallel}{\vert}^{2}
 +{\vert}H_{\perp}{\vert}^{2} \Big\}
   \label{br-v},
   \end{equation}
  %-----------------------------------------------------
  where ${\tau}_{B_{u}}$ $=$ $(1.638{\pm}0.004)$ ps is the
  lifetime of the $B_{u}$ meson \cite{pdg}.

  %-----------------------------------------------------
  \begin{table}[ht]
  \caption{The numerical values of the input parameters.}
  \label{tab:input}
  \begin{ruledtabular}
  \begin{tabular}{lll}
    CKM parameters\footnotemark[5]
  & $A$ $=$ $0.811{\pm}0.026$ \cite{pdg},
  & ${\lambda}$ $=$ $0.22506{\pm}0.00050$ \cite{pdg},  \\
  & $\bar{\rho}$ $=$ $0.124^{+0.019}_{-0.018}$ \cite{pdg},
  & $\bar{\eta}$ $=$ $0.356{\pm}0.011$ \cite{pdg}, \\ \hline
    mass of the particles
  & $m_{{\pi}^{\pm}}$ $=$ $139.57$ MeV \cite{pdg},
  & $m_{K^{\pm}}$ $=$ $493.677{\pm}0.016$ MeV \cite{pdg}, \\
  & $m_{\rho}$ $=$ $775.26{\pm}0.25$ MeV \cite{pdg},
  & $m_{K^{{\ast}{\pm}}}$ $=$ $891.66{\pm}0.26$ MeV \cite{pdg}, \\
    $m_{B_{u}}$ $=$ $5279.31{\pm}0.15$ MeV \cite{pdg},
  & $m_{b}$ $=$ $4.78{\pm}0.06$ GeV \cite{pdg},
  & $m_{c}$ $=$ $1.67{\pm}0.07$ GeV \cite{pdg}, \\ \hline
    decay constants
  & $f_{\pi}$ $=$ $130.2{\pm}1.7$ MeV \cite{pdg},
  & $f_{K}$ $=$ $155.6{\pm}0.4$ MeV \cite{pdg}, \\
  & $f_{\rho}$ $=$ $216{\pm}3$ MeV \cite{jhep0703.069},
  & $f_{K^{\ast}}$ $=$ $220{\pm}5$ MeV \cite{jhep0703.069}, \\
    $f_{B_{u}}$ $=$ $187.1{\pm}4.2$ MeV \cite{pdg},
  & $f_{\rho}^{T}$ $=$ $165{\pm}9$ MeV \cite{jhep0703.069},
  & $f_{K^{\ast}}^{T}$ $=$ $185{\pm}10$ MeV \cite{jhep0703.069}, \\ \hline
    \multicolumn{3}{l}{Gegenbauer moments at the scale of ${\mu}$ $=$ 1 GeV}\\
    $a_{1}^{K}$ $=$ $-0.06{\pm}0.03$ \cite{jhep0605.004},
  & $a_{2}^{K}$ $=$ $0.25{\pm}0.15$ \cite{jhep0605.004},
  & $a_{2}^{\pi}$ $=$ $0.25{\pm}0.15$ \cite{jhep0605.004}, \\
    $a_{1}^{{\parallel},K^{\ast}}$ $=$ $-0.03{\pm}0.02$ \cite{jhep0703.069},
  & $a_{2}^{{\parallel},K^{\ast}}$ $=$ $0.11{\pm}0.09$ \cite{jhep0703.069},
  & $a_{2}^{{\parallel},{\rho}}$ $=$ $0.15{\pm}0.07$ \cite{jhep0703.069}, \\
    $a_{1}^{{\perp},K^{\ast}}$ $=$ $-0.04{\pm}0.03$ \cite{jhep0703.069},
  & $a_{2}^{{\perp},K^{\ast}}$ $=$ $0.10{\pm}0.08$ \cite{jhep0703.069},
  & $a_{2}^{{\perp},{\rho}}$ $=$ $0.14{\pm}0.06$ \cite{jhep0703.069},
  \end{tabular}
  \end{ruledtabular}
  \footnotetext[5]{The relation between the CKM parameters (${\rho}$, ${\eta}$)
   and ($\bar{\rho}$, $\bar{\eta}$) is (${\rho}$, ${\eta}$) ${\simeq}$
   $(\bar{\rho}, \bar{\eta})(1+{\lambda}^{2}/2+{\cdots})$.}
  \end{table}
  %-----------------------------------------------------

  The numerical values of the input parameters are listed in Tables
  \ref{tab:psi} and \ref{tab:input}, where their central values will
  be regarded as the default inputs unless otherwise specified.
  Our numerical results for the branching ratios together with the
  experimental data are presented in Tables \ref{tab:br-1} and \ref{tab:br-2}.
  The theoretical uncertainties come from the quark mass $m_{c}$
  and $m_{b}$, and the hadronic parameters (including the decay constants,
  Gegenbauer moments, and the chiral parameter), respectively.
  The following are some comments.

  %-----------------------------------------------------
  \begin{table}[ht]
  \caption{The branching ratios for the $B_{u}$ ${\to}$
  ${\psi}(2S)M$, ${\psi}(3770)M$ decays, where the theoretical
  uncertainties come from the quark mass $m_{c}$, $m_{b}$,
  and the hadronic parameters, respectively.
  The numbers in the parentheses are the results without the
  nonfactorizable contributions.}
  \label{tab:br-1}
  \begin{ruledtabular}
  \begin{tabular}{cccccc}
    final states & unit & data \cite{pdg}
  & ${\theta}_{1}$ $=$ $0$
  & ${\theta}_{1}$ $=$ $-12^{\circ}$
  & ${\theta}_{1}$ $=$ $+27^{\circ}$ \\ \hline
    ${\psi}(2S)K^{-}$      & $10^{-4}$ & $6.26{\pm}0.24$
  & $11.77^{+ 0.22+ 1.92+ 4.99}_{- 0.24- 1.59- 3.88}$
  & $ 9.67^{+ 0.18+ 1.57+ 4.18}_{- 0.20- 1.29- 3.22}$
  & $12.93^{+ 0.24+ 2.12+ 5.65}_{- 0.26- 1.79- 4.35}$ \\ & &
  & ($13.24^{+ 0.00+ 2.08+ 5.38}_{- 0.00- 1.73- 4.22}$)
  & ($10.87^{+ 0.00+ 1.70+ 4.51}_{- 0.00- 1.40- 3.51}$)
  & ($14.56^{+ 0.00+ 2.30+ 6.11}_{- 0.00- 1.95- 4.75}$) \\ \hline
    ${\psi}(2S){\pi}^{-}$  & $10^{-5}$ & $2.44{\pm}0.30$
  & $ 1.91^{+ 0.05+ 0.42+ 0.87}_{- 0.05- 0.34- 0.66}$
  & $ 1.56^{+ 0.04+ 0.35+ 0.72}_{- 0.04- 0.28- 0.55}$
  & $ 2.10^{+ 0.05+ 0.46+ 0.99}_{- 0.06- 0.37- 0.74}$ \\ & &
  & ($ 2.13^{+ 0.00+ 0.46+ 0.92}_{- 0.00- 0.37- 0.71}$)
  & ($ 1.74^{+ 0.00+ 0.37+ 0.77}_{- 0.00- 0.30- 0.59}$)
  & ($ 2.35^{+ 0.00+ 0.50+ 1.06}_{- 0.00- 0.41- 0.80}$) \\ \hline
    ${\psi}(3770)K^{-}$      & $10^{-4}$ & $4.9{\pm}1.3$
  & $ 1.34^{+ 0.03+ 0.23+ 0.69}_{- 0.03- 0.21- 0.50}$
  & $ 3.33^{+ 0.06+ 0.56+ 1.60}_{- 0.07- 0.50- 1.19}$
  & $ 0.24^{+ 0.00+ 0.04+ 0.16}_{- 0.01- 0.02- 0.10}$ \\ & &
  & ($ 1.51^{+ 0.00+ 0.25+ 0.75}_{- 0.00- 0.23- 0.55}$)
  & ($ 3.77^{+ 0.00+ 0.61+ 1.74}_{- 0.00- 0.55- 1.31}$)
  & ($ 0.26^{+ 0.00+ 0.05+ 0.17}_{- 0.00- 0.02- 0.11}$) \\ \hline
    ${\psi}(3770){\pi}^{-}$  & $10^{-6}$ & ---
  & $ 2.24^{+ 0.06+ 0.49+ 1.26}_{- 0.07- 0.39- 0.88}$
  & $ 5.53^{+ 0.14+ 1.22+ 2.87}_{- 0.16- 0.97- 2.08}$
  & $ 0.37^{+ 0.01+ 0.08+ 0.25}_{- 0.01- 0.07- 0.16}$ \\ & &
  & ($ 2.50^{+ 0.00+ 0.54+ 1.35}_{- 0.00- 0.43- 0.96}$)
  & ($ 6.17^{+ 0.00+ 1.32+ 3.08}_{- 0.00- 1.06- 2.26}$)
  & ($ 0.40^{+ 0.00+ 0.09+ 0.27}_{- 0.00- 0.07- 0.17}$) \\ \hline
    ${\psi}(2S)K^{{\ast}-}$  & $10^{-4}$ & $6.7{\pm}1.4$
  & $12.88^{+ 0.44+ 2.11+ 4.05}_{- 0.63- 2.01- 3.31}$
  & $10.72^{+ 0.35+ 1.61+ 3.43}_{- 0.54- 1.75- 2.79}$
  & $13.80^{+ 0.51+ 2.60+ 4.52}_{- 0.65- 1.94- 3.65}$ \\ & &
  & ($ 9.76^{+ 0.00+ 1.91+ 3.14}_{- 0.00- 1.78- 2.47}$)
  & ($ 8.12^{+ 0.00+ 1.47+ 2.66}_{- 0.00- 1.55- 2.08}$)
  & ($10.47^{+ 0.00+ 2.36+ 3.51}_{- 0.00- 1.72- 2.73}$) \\ \hline
    ${\psi}(2S){\rho}^{-}$   & $10^{-5}$ & ---
  & $ 4.46^{+ 0.25+ 0.87+ 1.06}_{- 0.21- 0.75- 0.91}$
  & $ 3.67^{+ 0.21+ 0.71+ 0.89}_{- 0.17- 0.62- 0.76}$
  & $ 4.89^{+ 0.26+ 0.96+ 1.24}_{- 0.23- 0.83- 1.04}$ \\ & &
  & ($ 3.43^{+ 0.00+ 0.80+ 0.79}_{- 0.00- 0.67- 0.69}$)
  & ($ 2.81^{+ 0.00+ 0.65+ 0.66}_{- 0.00- 0.55- 0.58}$)
  & ($ 3.79^{+ 0.00+ 0.88+ 0.93}_{- 0.00- 0.75- 0.80}$) \\ \hline
    ${\psi}(3770)K^{{\ast}-}$  & $10^{-4}$ & ---
  & $ 1.20^{+ 0.06+ 0.42+ 0.50}_{- 0.04- 0.05- 0.37}$
  & $ 3.21^{+ 0.15+ 0.88+ 1.20}_{- 0.13- 0.29- 0.93}$
  & $ 0.36^{+ 0.00+ 0.00+ 0.16}_{- 0.03- 0.12- 0.12}$ \\ & &
  & ($ 0.92^{+ 0.00+ 0.38+ 0.39}_{- 0.00- 0.05- 0.28}$)
  & ($ 2.45^{+ 0.00+ 0.79+ 0.93}_{- 0.00- 0.26- 0.70}$)
  & ($ 0.26^{+ 0.00+ 0.00+ 0.12}_{- 0.00- 0.10- 0.09}$) \\ \hline
    ${\psi}(3770){\rho}^{-}$   & $10^{-6}$ & ---
  & $ 4.93^{+ 0.17+ 1.02+ 1.69}_{- 0.20- 0.87- 1.30}$
  & $12.35^{+ 0.53+ 2.49+ 3.74}_{- 0.53- 2.14- 2.99}$
  & $ 0.92^{+ 0.10+ 0.16+ 0.38}_{- 0.05- 0.14- 0.30}$ \\ & &
  & ($ 3.95^{+ 0.00+ 0.95+ 1.32}_{- 0.00- 0.79- 1.02}$)
  & ($ 9.76^{+ 0.00+ 2.32+ 2.87}_{- 0.00- 1.94- 2.31}$)
  & ($ 0.65^{+ 0.00+ 0.14+ 0.26}_{- 0.00- 0.12- 0.20}$)
  \end{tabular}
  \end{ruledtabular}
  \end{table}
  %-----------------------------------------------------
  \begin{table}[ht]
  \caption{The branching ratios for the $B_{u}$ ${\to}$
  ${\psi}(4040)M$, ${\psi}(4160)M$ decays, where the theoretical
  uncertainties come from the quark mass $m_{c}$, $m_{b}$,
  and the hadronic parameters, respectively.
  The numbers in the parentheses are the results without
  the nonfactorizable contributions.}
  \label{tab:br-2}
  \begin{ruledtabular}
  \begin{tabular}{ccccc}
    final states & unit & data \cite{pdg}
  & ${\theta}_{2}$ $=$ $0$
  & ${\theta}_{2}$ $=$ $-35^{\circ}$ \\ \hline
    ${\psi}(4040)K^{-}$      & $10^{-4}$ & $<1.3$
  & $ 4.21^{+ 0.06+ 0.77+ 2.55}_{- 0.05- 0.58- 1.77}$
  & $ 0.52^{+ 0.01+ 0.09+ 1.07}_{- 0.01- 0.07- 0.41}$ \\ & &
  & ($ 5.08^{+ 0.00+ 0.86+ 2.91}_{- 0.00- 0.66- 2.06}$)
  & ($ 0.61^{+ 0.00+ 0.10+ 1.24}_{- 0.00- 0.08- 0.49}$) \\ \hline
    ${\psi}(4040){\pi}^{-}$  & $10^{-6}$ & ---
  & $ 7.84^{+ 0.21+ 1.77+ 5.16}_{- 0.17- 1.30- 3.49}$
  & $ 0.85^{+ 0.03+ 0.21+ 2.00}_{- 0.02- 0.16- 0.71}$ \\ & &
  & ($ 9.47^{+ 0.00+ 1.99+ 6.02}_{- 0.00- 1.47- 4.10}$)
  & ($ 1.00^{+ 0.00+ 0.23+ 2.36}_{- 0.00- 0.18- 0.84}$) \\  \hline
    ${\psi}(4160)K^{-}$      & $10^{-4}$ & $5.1{\pm}2.7$
  & $ 2.21^{+ 0.03+ 0.42+ 3.07}_{- 0.02- 0.30- 1.47}$
  & $ 5.41^{+ 0.08+ 1.02+ 5.36}_{- 0.06- 0.74- 3.00}$ \\ & &
  & ($ 2.72^{+ 0.00+ 0.48+ 3.63}_{- 0.00- 0.34- 1.78}$)
  & ($ 6.59^{+ 0.00+ 1.14+ 6.26}_{- 0.00- 0.84- 3.57}$) \\  \hline
    ${\psi}(4160){\pi}^{-}$  & $10^{-6}$ & ---
  & $ 4.65^{+ 0.12+ 1.00+ 6.87}_{- 0.09- 0.70- 3.15}$
  & $10.88^{+ 0.28+ 2.40+11.64}_{- 0.22- 1.72- 6.26}$ \\ & &
  & ($ 5.71^{+ 0.00+ 1.13+ 8.16}_{- 0.00- 0.80- 3.83}$)
  & ($13.25^{+ 0.00+ 2.70+13.75}_{- 0.00- 1.95- 7.51}$) \\  \hline
    ${\psi}(4040)K^{{\ast}-}$  & $10^{-4}$ & ---
  & $ 2.97^{+ 0.03+ 0.71+ 1.61}_{- 0.05- 0.62- 1.10}$
  & $ 0.75^{+ 0.00+ 0.23+ 0.80}_{- 0.01- 0.20- 0.39}$ \\ & &
  & ($ 2.17^{+ 0.00+ 0.62+ 1.26}_{- 0.00- 0.53- 0.84}$)
  & ($ 0.58^{+ 0.00+ 0.21+ 0.61}_{- 0.00- 0.18- 0.29}$) \\  \hline
    ${\psi}(4040){\rho}^{-}$   & $10^{-5}$ & ---
  & $ 1.52^{+ 0.02+ 0.16+ 0.64}_{- 0.02- 0.28- 0.47}$
  & $ 0.26^{+ 0.00+ 0.02+ 0.33}_{- 0.00- 0.14- 0.16}$ \\ & &
  & ($ 1.19^{+ 0.00+ 0.15+ 0.53}_{- 0.00- 0.26- 0.38}$)
  & ($ 0.21^{+ 0.00+ 0.02+ 0.26}_{- 0.00- 0.12- 0.13}$) \\  \hline
    ${\psi}(4160)K^{{\ast}-}$  & $10^{-4}$ & ---
  & $ 0.69^{+ 0.01+ 0.37+ 0.92}_{- 0.01- 0.28- 0.43}$
  & $ 2.32^{+ 0.03+ 0.87+ 2.05}_{- 0.04- 0.67- 1.15}$ \\ & &
  & ($ 0.47^{+ 0.00+ 0.31+ 0.68}_{- 0.00- 0.22- 0.30}$)
  & ($ 1.63^{+ 0.00+ 0.76+ 1.55}_{- 0.00- 0.57- 0.84}$) \\  \hline
    ${\psi}(4160){\rho}^{-}$   & $10^{-5}$ & ---
  & $ 0.56^{+ 0.01+ 0.15+ 0.66}_{- 0.00- 0.11- 0.34}$
  & $ 1.57^{+ 0.02+ 0.18+ 1.21}_{- 0.01- 0.30- 0.73}$ \\ & &
  & ($ 0.43^{+ 0.00+ 0.13+ 0.54}_{- 0.00- 0.10- 0.27}$)
  & ($ 1.21^{+ 0.00+ 0.16+ 0.99}_{- 0.00- 0.26- 0.58}$)
  \end{tabular}
  \end{ruledtabular}
  \end{table}
  %-----------------------------------------------------

  (1)
  It has been shown in Refs.\cite{npb606.245,plb509.263,prd64.014036}
  that the contributions from the spectator scattering topologies
  to the coefficient $a_{2}$ with the QCDF approach are amplified
  by the large Wilson coefficient $C_{1}$, and the contributions are
  notable for the $B$ ${\to}$ $J/{\psi}M$ decays \cite{prd78.074030,prd63.074011}.
  Hence, it is initially expected that the nonfactorizable contributions
  from Fig.\ref{fig:fey}(c,d) should be significant for the
  $B_{u}$ ${\to}$ ${\psi}M$ decays.
  However, it is seen from the numbers in Tables \ref{tab:br-1} and
  \ref{tab:br-2} that compared with the factorizable contributions,
  the nonfactorizable contributions to the branching ratios are important,
  but not so obvious as expected.
  One of the reasons might be that the opposite signs of
  the charm quark propagators of Fig.\ref{fig:fey}(c) and Fig.\ref{fig:fey}(d)
  results in the destructive interference between their amplitudes.
  In addition, the amplitudes of Fig.\ref{fig:fey}(c,d) are suppressed
  by the color factor $1/N_{c}$ relative to the amplitudes of Fig.\ref{fig:fey}(a,b)
  (see the expressions listed in Appendix \ref{block}).
  It is also shown that the nonfactorizable contributions are
  positive (negative) to branching ratios for the $B_{u}$ ${\to}$ ${\psi}V$
  (${\psi}P$) decays.

  (2)
  The $B_{u}$ ${\to}$ ${\psi}(2S)M$ decays have been studied with the
  pQCD approach in Refs.\cite{epjc77.610,plb772.719}, by considering
  part of the NLO factorizable vertex corrections,
  but without the $2S$-$1D$ mixing effects on psions.
  Our numerical results generally agree with those of Refs.\cite{epjc77.610,plb772.719}
  within theoretical uncertainties, although with different parameters.
  In the future, a careful and comprehensive study of the NLO corrections to
  the $B$ ${\to}$ ${\psi}M$ decays is desperately needed, and will be
  essential for the forthcoming precision measurements at the LHCb and Belle-II
  experiments.

  %-----------------------------------------------------
  \begin{figure}[h]
  \includegraphics[width=0.95\textwidth,bb=75 510 530 720]{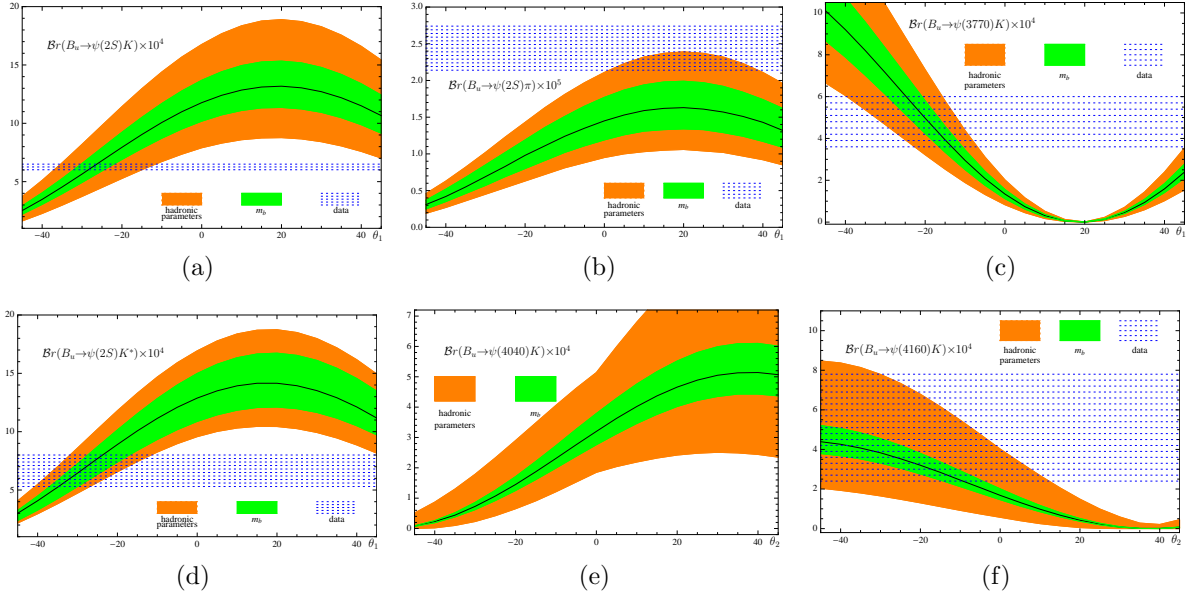}
  \caption{The branching ratios (vertical axis) versus the
  $S$-$D$ mixing angle (horizontal axis, in degrees).
  The solid lines denote the results calculated with the
  default inputs; the dotted blocks denote the current experimental
  data within one standard error; the green and orange blocks
  correspond to theoretical uncertainties from of
  $m_{b}$ and hadronic parameters, respectively.}
  \label{fig:phi}
  \end{figure}
  %-----------------------------------------------------

  (3)
  The $S$-$D$ wave mixture has literally altered the branching
  ratios for the $B_{u}$ ${\to}$ ${\psi}M$ decays.
  The $B_{u}$ ${\to}$ ${\psi}(2S)K^{(\ast)}$, ${\psi}(3770)K$,
  ${\psi}(4160)K$, ${\psi}(2S){\pi}$ decays can be reasonably
  accommodated within theoretical uncertainties with the appropriate
  $S$-$D$ wave mixing angles and other inputs.
  The angle ${\theta}_{1}$ for $2S$-$1D$ mixing and
  ${\theta}_{2}$ for $3S$-$2D$ mixing prefer the negative values,
  except for the $B_{u}$ ${\to}$ ${\psi}(2S){\pi}$ decay.
  However, the current experimental data for the $B_{u}$ ${\to}$ ${\psi}M$
  decays cannot offer the $S$-$D$ mixing angles (${\theta}_{1}$
  and ${\theta}_{2}$) with a severe constraint (also see Fig.\ref{fig:phi}).
  With the successful implementation of the high-luminosity LHCb
  and SuperKEKB experiments,
  more accurate measurements of the $B$ ${\to}$ ${\psi}M$ decays will
  be obtained. In addition, a comprehensive study with more processes
  pertinent to the psions, including the pure leptonic psion decays
  and the $B$ ${\to}$ ${\psi}M$ decays, are necessary to determine
  the $S$-$D$ wave mixing angles in the future.

  %-----------------------------------------------------
  \begin{figure}[h]
  \includegraphics[width=0.5\textwidth,bb=110 470 500 730]{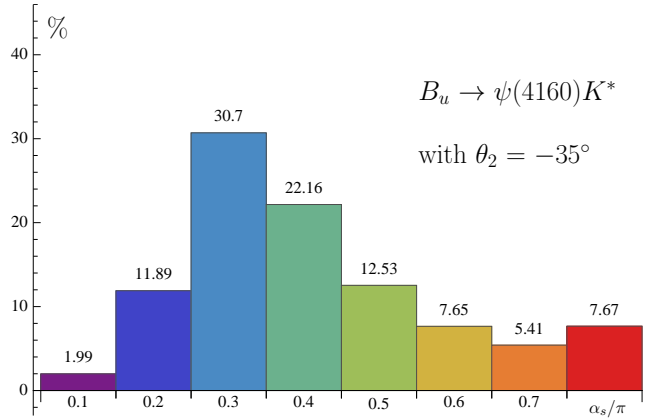}
  \caption{The percentage contribution to branching ratio for the
  $B_{u}$ ${\to}$ ${\psi}(4160)K^{\ast}$ decay versus ${\alpha}_{s}/{\pi}$,
  where the numbers above the histogram denote
  the percentage.}
  \label{fig:his}
  \end{figure}
  %-----------------------------------------------------

  (4)
  The excited psions with a large mass will certainly  carry a
  large portion of energy in the $B$ meson decay when they are emitted
  from the interaction point. It is therefore natural to doubt whether the
  gluons exchanged between the $B$ meson and the recoiled $M$ meson
  are hard enough to validate the perturbative calculation
  and the practicability of the pQCD approach.
  In addition, it is shown in Ref.\cite{prd71.114008} that the variation
  of the renormalization scale has a great impact on the color-suppressed
  $B$ ${\to}$ $J/{\psi}M$, ${\eta}_{c}M$ decays. In order to clear these
  doubts, it is necessary to check how many shares come from the
  perturbative domain. The ${\psi}(4160)$ meson has the largest
  mass among the psions concerned.
  To make the analysis more persuasive, we take the $B_{u}$ ${\to}$
  ${\psi}(4160)K^{\ast}$ decay as an example.
  The percentage contributions to the branching ratio from different ${\alpha}_{s}/{\pi}$
  regions are shown in Fig.\ref{fig:his}.
  It is seen that more than 60\% of contributions come from the
  ${\alpha}_{s}/{\pi}$ ${\le}$ $0.4$ regions. Our study also shows that
  more than 80\% of contributions to the $B_{u}$ ${\to}$ ${\psi}(2S){\pi}$
  decay come from the ${\alpha}_{s}/{\pi}$ ${\le}$ $0.4$ regions. These facts
  imply that the perturbative calculation with the pQCD approach might be
  feasible. Besides the suppression on the soft contributions from both the
  Sudakov factors and the exponential functions of DAs in
  Eqs.(\ref{wave-bup}-\ref{wave-v-T}),
  the choice of the renormalization scale as the maximum among all possible virtualities
  [see Eq.(\ref{scale-ti})] is also an important factor to further ensure
  the perturbative calculation with the pQCD approach.

  (5)
  Because of the large mass of the excited psions, the phase space
  for the $B_{u}$ ${\to}$ ${\psi}M$ decays is relatively compact.
  For example, the total kinetic energy of the final states for the $B_{u}$
  ${\to}$ ${\psi}(4160)K^{\ast}$ decay is $m_{B_{u}}$ $-$
  $m_{\psi}$ $-$ $m_{M}$ $<$ 200 MeV.
  Hence, the final state interactions (FSIs) might have a non-negligible
  influence on the $B_{u}$ ${\to}$ ${\psi}M$ decays.
  Overlooking FSIs might be one reason why the QCDF approach is not
  good enough for the $B$ ${\to}$ $J/{\psi}M$ decays in
  Refs.\cite{prd63.074011,prd65.094023}.
  The potential FSIs deserve much attention for the
  nonleptonic $B$ ${\to}$ ${\psi}M$ decays, but this is
  beyond the scope of this paper.

  (6)
  There are lots of theoretical uncertainties, especially from $m_{b}$,
  hadronic parameters and the $S$-$D$ mixing angles.
  It is shown in Refs.\cite{cpc34.937,prd89.094010,prd71.114008,epjc77.610}
  that the pQCD's results are sensitive to the model of mesonic WFs/DAs
  and input parameters.
  Besides, many other factors, such as FSIs, different models for
  mesonic WFs/DAs, higher order corrections to HME, and so on,
  are not scrutinized here, in spite of the value of dedicated study.
  Most of the theoretical uncertainties actually result from our
  inadequate comprehension of the long-distance and nonperturbative dynamics.
  Great efforts should be made to improve the reliability of theoretical
  results.

  %%%%%%%%%%%%%%%%%%%%%%%%%%%%%%%%%%%%%%%%%%%%%%%%%%%%%%%%%%%
  \section{Summary}
  \label{sec04}
  The color-suppressed nonleptonic $B_{u}$ ${\to}$ ${\psi}M$ decay
  provides an important place to explore the $S$-$D$ wave mixing among
  psions, and test the QCD-inspired approaches for dealing with the
  hadronic matrix elements.
  In this paper, the $B_{u}$ ${\to}$ ${\psi}M$ decays are
  investigated with the pQCD approach, including the contributions
  of factorizable and nonfactorizable emission topologies.
  We also consider the effects of $2S$-$1D$ and $3S$-$2D$ mixing
  on psions.
  It is found that with appropriate inputs, there is generally agreement
  with the experimental data for the branching ratios for the $B_{u}$
  ${\to}$ ${\psi}K$ decays within theoretical uncertainties.
  However, due to the large experimental and theoretical uncertainties,
  the angle ${\theta}_{1}$ (${\theta}_{2}$) for the $2S$-$1D$ ($3S$-$2D$)
  wave mixing cannot be determined properly for the moment.

  %%%%%%%%%%%%%%%%%%%%%%%%%%%%%%%%%%%%%%%%%%%%%%%%%%%%%%%%%%%
  \section*{Acknowledgments}
  The work is supported by the National Natural Science Foundation
  of China (Grant Nos. 11705047, U1632109, 11547014 and 11475055),
  and Open Research Program of Large Research Infrastructures
  (2017), Chinese Academy of Sciences. We thank Ms. Nan Li (HNU)
  for polishing this paper.

  %%%%%%%%%%%%%%%%%%%%%%%%%%%%%%%%%%%%%%%%%%%%%%%%%%%%%%%%%%%
  \begin{appendix}
  \section{Wave functions for the $nS$ and $nD$ charmonium states}
  \label{cc-wfs}
  The charmonium systems
  are usually assumed to be nonrelativistic, and their wave functions
  can be obtained from the solutions of the time-independent Schr\"{o}dinger equation.
  Here, we will take the conventional notation to specify the ${\psi}(nL)$
  states, where $n$ $=$ $1$, $2$, $3$, ${\cdots}$ is the radial quantum number,
  and the orbital angular momentum $L$ $=$ $0$, $1$, $2$ ${\cdots}$ corresponds
  to $S$, $P$, $D$ ${\cdots}$ waves, respectively.
  The wave functions for the $nS$ and $nD$ states associated with the isotropic
  linear harmonic oscillator potential are written as follows.
  %-----------------------------------------------------
  \begin{eqnarray}
 {\psi}_{1S}(\vec{k}) &{\sim}& e^{-\frac{\vec{k}^{2}}{2\,{\omega}^{2}}}
  \label{p-wf-1s}, \\
 {\psi}_{2S}(\vec{k}) &{\sim}& e^{-\frac{\vec{k}^{2}}{2\,{\omega}^{2}}}\,
         (2\,\vec{k}^{2}-3\,{\omega}^{2})
  \label{p-wf-2s}, \\
 {\psi}_{3S}(\vec{k}) &{\sim}& e^{-\frac{\vec{k}^{2}}{2\,{\omega}^{2}}}\,
         (4\,\vec{k}^{4}-20\,\vec{k}^{2}\,{\omega}^{2}-15\,{\omega}^{4})
  \label{p-wf-3s}, \\
 {\psi}_{1D}(\vec{k}) &{\sim}& \vec{k}^{2}\,e^{-\frac{\vec{k}^{2}}{2\,{\omega}^{2}}}
  \label{p-wf-1d}, \\
 {\psi}_{2D}(\vec{k}) &{\sim}& \vec{k}^{2}\,e^{-\frac{\vec{k}^{2}}{2\,{\omega}^{2}}}\,
         (2\,\vec{k}^{2}-7\,{\omega}^{2})
  \label{p-wf-2d},
  \end{eqnarray}
  %-----------------------------------------------------
  where the parameter ${\omega}$ determines the average transverse momentum of
  the oscillator, i.e., ${\langle}1S{\vert}\vec{k}^{2}_{T}{\vert}1S{\rangle}$
  $=$ ${\omega}^{2}$.
  With the power counting rules of the nonrelativistic QCD effective theory
  \cite{prd46.4052,prd51.1125,rmp77.1423},
  the characteristic velocity $v$ of the valence quark in heavy quarkonium
  is about $v$ ${\sim}$ ${\alpha}_{s}$.
  The parameter ${\omega}$ ${\simeq}$ $m\,{\alpha}_{s}$
  is taken for the psions in our calculation, where ${\alpha}_{s}$ is
  the QCD coupling constant. We adopt the light-cone momentum
  and employ the commonly used substitution \cite{epja15.523},
  %-----------------------------------------------------
   \begin{equation}
   \vec{k}^{2}\ {\to}\ \frac{1}{4} \sum\limits_{i}
   \frac{\vec{k}_{iT}^{2}+m_{q_{i}}^{2}}{x_{i}}
   \label{wave-kt},
   \end{equation}
  %-----------------------------------------------------
  where $x_{i}$, $\vec{k}_{iT}$, $m_{q_{i}}$ are the longitudinal momentum
  fraction, transverse momentum, and mass of the valence quark.
  These variables satisfy the
  relations ${\sum}x_{i}$ $=$ $1$ and $\sum\vec{k}_{iT}$ $=$ $0$.
  After integrating out $\vec{k}_{iT}$ and combining the results with their asymptotic
  forms \cite{jhep9901.010,jhep0703.069,jhep0605.004}, one can obtain the distribution
  amplitudes of Eqs.(\ref{wave-1s-v}-\ref{wave-v-T}) for the charmonium states.

  \section{Amplitude building blocks for the $B_{u}^{-}$ ${\to}$ ${\psi}M$ decays}
  \label{block}
  %-----------------------------------------------------
  %------------------------------------
   \begin{eqnarray}
  {\cal A}^{LL,LR}_{a,P} &=&
  {\int}_{0}^{1}{\rm d}x_{1}
  {\int}_{0}^{1}{\rm d}x_{3}
  {\int}_{0}^{\infty}b_{1}\,{\rm d}b_{1}
  {\int}_{0}^{\infty}b_{3}\,{\rm d}b_{3}\,
  H_{f}({\alpha},{\beta}_{a},b_{1},b_{3})\,
  E_{f}(t_{a})\, {\alpha}_{s}(t_{a})
   \nonumber \\ &{\times}&
   \Big\{ {\phi}_{B}^{a}(x_{1})\,\Big[
  2\,m_{1}\,p\,\{ {\phi}_{P}^{a}(x_{3})\,
  (m_{1}^{2}\,\bar{x}_{3}+m_{2}^{2}\,x_{3})
  +m_{b}\,{\mu}_{P}\,{\phi}_{P}^{p}(x_{3}) \}
  \nonumber \\ & & +\,
   t\,m_{b}\,{\mu}_{P}\,{\phi}_{P}^{t}(x_{3}) \Big]
  -2\,m_{1}\,{\phi}_{B}^{p}(x_{1})\,\Big[
   2\,m_{1}\,p\,m_{b}\,{\phi}_{P}^{a}(x_{3})\,
   \nonumber \\ & &+\,
   2\,m_{1}\,p\,{\mu}_{P}\,{\phi}_{P}^{p}(x_{3})\,\bar{x}_{3}
  +{\mu}_{P}\,{\phi}_{P}^{t}(x_{3})\,(t-s\,x_{3})
   \Big] \Big\}
   \label{amp:a-p},
   \end{eqnarray}
  %------------------------------------
  %------------------------------------
   \begin{eqnarray}
  {\cal A}^{LL,LR}_{a,L} &=&
  {\int}_{0}^{1}{\rm d}x_{1}
  {\int}_{0}^{1}{\rm d}x_{3}
  {\int}_{0}^{\infty}b_{1}\,{\rm d}b_{1}
  {\int}_{0}^{\infty}b_{3}\,{\rm d}b_{3}\,
  H_{f}({\alpha},{\beta}_{a},b_{1},b_{3})\,
  E_{f}(t_{a})\, {\alpha}_{s}(t_{a})
   \nonumber \\ &{\times}&
   \Big\{ 2\,m_{1}\,{\phi}_{B}^{p}(x_{1})\,
   \Big[ {\phi}_{V}^{s}(x_{3})\,2\,m_{1}\,m_{3}\,p\,\bar{x}_{3}
 +{\phi}_{V}^{t}(x_{3})\,m_{3}\,(t-s\,x_{3})
   \nonumber \\ & &+\,
  {\phi}_{V}^{v}(x_{3})\,m_{b}\,s
   \Big]-{\phi}_{B}^{a}(x_{1})\, \Big[
  {\phi}_{V}^{v}(x_{3})\,(m_{1}^{2}\,s\,\bar{x}_{3}
  +m_{2}^{2}\,u\,x_{3})
   \nonumber \\ & &+\,
  {\phi}_{V}^{t}(x_{3})\,m_{3}\,m_{b}\,t
 +{\phi}_{V}^{s}(x_{3})\,2\,m_{1}\,p\,m_{3}\,m_{b}
   \Big] \Big\}
   \label{amp:a-l},
   \end{eqnarray}
  %------------------------------------
  %------------------------------------
   \begin{eqnarray}
  {\cal A}^{LL,LR}_{a,N} &=&
  {\int}_{0}^{1}{\rm d}x_{1}
  {\int}_{0}^{1}{\rm d}x_{3}
  {\int}_{0}^{\infty}b_{1}\,{\rm d}b_{1}
  {\int}_{0}^{\infty}b_{3}\,{\rm d}b_{3}\,
  H_{f}({\alpha},{\beta}_{a},b_{1},b_{3})\,
  E_{f}(t_{a})\, {\alpha}_{s}(t_{a})
   \nonumber \\ &{\times}&
   \Big\{ {\phi}_{B}^{p}(x_{1})\,2\,m_{1}\,m_{2}\,
   \Big[ {\phi}_{V}^{V}(x_{3})\,2\,m_{3}\,m_{b}
 +{\phi}_{V}^{T}(x_{3})\,(u-2\,m_{3}^{2}\,x_{3}) \Big]
   \nonumber \\ & & -\,
  {\phi}_{B}^{a}(x_{1})\,\Big[
  {\phi}_{V}^{V}(x_{3})\,m_{2}\,m_{3}\,(2\,m_{1}^{2}-u\,x_{3})
 +{\phi}_{V}^{T}(x_{3})\,m_{2}\,m_{b}\,u
   \nonumber \\ & & +\,
  {\phi}_{V}^{A}(x_{3})\,2\,m_{1}\,m_{2}\,m_{3}\,p\,x_{3}
   \Big] \Big\}
   \label{amp:a-n},
   \end{eqnarray}
  %------------------------------------
  %------------------------------------
   \begin{eqnarray}
  {\cal A}^{LL,LR}_{a,T} &=&
  {\int}_{0}^{1}{\rm d}x_{1}
  {\int}_{0}^{1}{\rm d}x_{3}
  {\int}_{0}^{\infty}b_{1}\,{\rm d}b_{1}
  {\int}_{0}^{\infty}b_{3}\,{\rm d}b_{3}\,
  H_{f}({\alpha},{\beta}_{a},b_{1},b_{3})\,
  E_{f}(t_{a})\, {\alpha}_{s}(t_{a})
   \nonumber \\ &{\times}&
  m_{2}\,\Big\{ {\phi}_{B}^{a}(x_{1})\,\Big[
  m_{3}/(m_{1}\,p)\,
  {\phi}_{V}^{A}(x_{3})\,(2\,m_{1}^{2}-u\,x_{3})
 +{\phi}_{V}^{T}(x_{3})\,2\,m_{b}
   \nonumber \\ &+&
  {\phi}_{V}^{V}(x_{3})\,2\,m_{3}\,x_{3}
   \Big] - 4\,{\phi}_{B}^{p}(x_{1})\, \Big[
  {\phi}_{V}^{T}(x_{3})\,m_{1}
 +{\phi}_{V}^{A}(x_{3})\,m_{3}\,m_{b}/p
   \Big] \Big\}
   \label{amp:a-t},
   \end{eqnarray}
  %------------------------------------
  %-----------------------------------------------------
  %------------------------------------
   \begin{eqnarray}
  {\cal A}^{LL,LR}_{b,P} &=& 2\,m_{1}\,p\,
  {\int}_{0}^{1}{\rm d}x_{1}
  {\int}_{0}^{1}{\rm d}x_{3}
  {\int}_{0}^{\infty}b_{1}\,{\rm d}b_{1}
  {\int}_{0}^{\infty}b_{3}\,{\rm d}b_{3}\,
  H_{f}({\alpha},{\beta}_{b},b_{3},b_{1})\,
  E_{f}(t_{b})\, {\alpha}_{s}(t_{b})
   \nonumber \\ &{\times}&
   \Big\{ {\phi}_{B}^{a}(x_{1})\, {\phi}_{P}^{a}(x_{3})\,
  (m_{3}^{2}\,\bar{x}_{1}+m_{2}^{2}\,x_{1})
  -{\phi}_{B}^{p}(x_{1})\, {\phi}_{P}^{p}(x_{3})\,
  2\,m_{1}\,{\mu}_{P}\,\bar{x}_{1} \Big\}
   \label{amp:b-p},
   \end{eqnarray}
  %------------------------------------
  %------------------------------------
   \begin{eqnarray}
  {\cal A}^{LL,LR}_{b,L} &=&
  {\int}_{0}^{1}{\rm d}x_{1}
  {\int}_{0}^{1}{\rm d}x_{3}
  {\int}_{0}^{\infty}b_{1}\,{\rm d}b_{1}
  {\int}_{0}^{\infty}b_{3}\,{\rm d}b_{3}\,
  H_{f}({\alpha},{\beta}_{b},b_{3},b_{1})\,
  E_{f}(t_{b})\, {\alpha}_{s}(t_{b})
   \nonumber \\ & & \!\!\!\! \!\!\!\! \!\!\!\! {\times}\,
   \Big\{ {\phi}_{B}^{a}(x_{1})\, {\phi}_{V}^{v}(x_{3})\,
   (m_{3}^{2}\,t\,\bar{x}_{1}-m_{2}^{2}\,u\,x_{1})
  +{\phi}_{B}^{p}(x_{1})\, {\phi}_{V}^{s}(x_{3})\,
  4\,m_{1}^{2}\,m_{3}\,p\,\bar{x}_{1} \Big\}
   \label{amp:b-l},
   \end{eqnarray}
  %------------------------------------
  %------------------------------------
   \begin{eqnarray}
  {\cal A}^{LL,LR}_{b,N} &=&
  m_{2}\,m_{3}\, {\int}_{0}^{1}{\rm d}x_{1}
  {\int}_{0}^{1}{\rm d}x_{3}
  {\int}_{0}^{\infty}b_{1}\,{\rm d}b_{1}
  {\int}_{0}^{\infty}b_{3}\,{\rm d}b_{3}\,
  H_{f}({\alpha},{\beta}_{b},b_{3},b_{1})\,
  E_{f}(t_{b})
   \nonumber \\ &{\times}&
  {\alpha}_{s}(t_{b})\,
  {\phi}_{B}^{a}(x_{1})\, \Big\{
  {\phi}_{V}^{V}(x_{3})\, (u-2\,m_{1}^{2}\,x_{1})
 +{\phi}_{V}^{A}(x_{3})\, 2\,m_{1}\,p \Big\}
   \label{amp:b-n},
   \end{eqnarray}
  %------------------------------------
  %------------------------------------
   \begin{eqnarray}
  {\cal A}^{LL,LR}_{b,T} &=&
  -m_{2}\,m_{3}\, {\int}_{0}^{1}{\rm d}x_{1}
  {\int}_{0}^{1}{\rm d}x_{3}
  {\int}_{0}^{\infty}b_{1}\,{\rm d}b_{1}
  {\int}_{0}^{\infty}b_{3}\,{\rm d}b_{3}\,
  H_{f}({\alpha},{\beta}_{b},b_{3},b_{1})\,
  E_{f}(t_{b})
   \nonumber \\ &{\times}&
  {\alpha}_{s}(t_{b})\,
  {\phi}_{B}^{a}(x_{1})\, \Big\{
  2\,{\phi}_{V}^{V}(x_{3})
 +{\phi}_{V}^{A}(x_{3})\,
  (u-2\,m_{1}^{2}\,x_{1})/(m_{1}\,p) \Big\}
   \label{amp:b-t},
   \end{eqnarray}
  %------------------------------------
  %-----------------------------------------------------
  %------------------------------------
   \begin{eqnarray}
  {\cal A}^{LL}_{c,P} &=&
   \frac{1}{N_{c}}\,
  {\int}_{0}^{1}{\rm d}x_{1}
  {\int}_{0}^{1}{\rm d}x_{2}
  {\int}_{0}^{1}{\rm d}x_{3}
  {\int}_{0}^{\infty}{\rm d}b_{1}
  {\int}_{0}^{\infty}b_{2}\,{\rm d}b_{2}
  {\int}_{0}^{\infty}b_{3}\,{\rm d}b_{3}\,
  H_{n}({\alpha},{\beta}_{c},b_{2},b_{3})
   \nonumber \\ &{\times}&
  E_{n}(t_{c})\, \Big\{
  {\phi}_{B}^{a}(x_{1})\, {\phi}_{P}^{a}(x_{3})\,
  2\, m_{1}\, p\, \Big[ {\phi}_{\psi}^{v}(x_{2})\,
  \{ u\,(x_{1}-x_{3})+s\,(x_{3}-\bar{x}_{2}) \}
  \nonumber \\ &-&
  {\phi}_{\psi}^{t}(x_{2})\, m_{2}\,m_{c} \Big] +
  {\phi}_{B}^{p}(x_{1})\, {\phi}_{\psi}^{v}(x_{2})\,
   m_{1}\,{\mu}_{P}\, \Big[ {\phi}_{P}^{p}(x_{3})\,
  2\,m_{1}\,p\,(x_{3}-x_{1})
   \nonumber \\ &+&
  {\phi}_{P}^{t}(x_{3})\, \{
  t\,(x_{1}-\bar{x}_{2})+s\,(\bar{x}_{2}-x_{3}) \}
   \Big] \Big\}\, {\alpha}_{s}(t_{c})\,
  {\delta}(b_{1}-b_{3})
   \label{amp:ll-c-p},
   \end{eqnarray}
  %------------------------------------
  %------------------------------------
   \begin{eqnarray}
  {\cal A}^{LL}_{c,L} &=&
   \frac{1}{N_{c}}\,
  {\int}_{0}^{1}{\rm d}x_{1}
  {\int}_{0}^{1}{\rm d}x_{2}
  {\int}_{0}^{1}{\rm d}x_{3}
  {\int}_{0}^{\infty}{\rm d}b_{1}
  {\int}_{0}^{\infty}b_{2}\,{\rm d}b_{2}
  {\int}_{0}^{\infty}b_{3}\,{\rm d}b_{3}\,
  H_{n}({\alpha},{\beta}_{c},b_{2},b_{3})
   \nonumber \\ &{\times}&
   \Big\{
  {\phi}_{B}^{a}(x_{1})\, {\phi}_{V}^{v}(x_{3})\,
   \Big[ {\phi}_{\psi}^{v}(x_{2})\,4\,m_{1}^{2}\,
  p^{2}\,(\bar{x}_{2}-x_{1})
 +{\phi}_{\psi}^{t}(x_{2})\,m_{2}\,m_{c}\,u \Big]
  \nonumber \\ &+&
 {\phi}_{B}^{p}(x_{1})\, {\phi}_{\psi}^{v}(x_{2})\,
  m_{1}\,m_{3}\, \Big[ {\phi}_{V}^{t}(x_{3})\,
  \{ t\,(\bar{x}_{2}-x_{1})+s\,(x_{3}-\bar{x}_{2}) \}
   \nonumber \\ &+&
  {\phi}_{V}^{s}(x_{3})\, 2\,m_{1}\,p\,(x_{1}-x_{3})
   \Big] \Big\}\, E_{n}(t_{c})\, {\alpha}_{s}(t_{c})\,
  {\delta}(b_{1}-b_{3})
   \label{amp:ll-c-l},
   \end{eqnarray}
  %------------------------------------
  %------------------------------------
   \begin{eqnarray}
  {\cal A}^{LL}_{c,N} &=&
   \frac{1}{N_{c}}\,
  {\int}_{0}^{1}{\rm d}x_{1}
  {\int}_{0}^{1}{\rm d}x_{2}
  {\int}_{0}^{1}{\rm d}x_{3}
  {\int}_{0}^{\infty}{\rm d}b_{1}
  {\int}_{0}^{\infty}b_{2}\,{\rm d}b_{2}
  {\int}_{0}^{\infty}b_{3}\,{\rm d}b_{3}\,
  H_{n}({\alpha},{\beta}_{c},b_{2},b_{3})
   \nonumber \\ &{\times}&
  E_{n}(t_{c})\, {\delta}(b_{1}-b_{3})\,
   \Big\{ {\phi}_{B}^{a}(x_{1})\,
  {\phi}_{\psi}^{T}(x_{2})\, m_{3}\,m_{c}\, \Big[
  {\phi}_{V}^{V}(x_{3})\,t
 -{\phi}_{V}^{A}(x_{3})\,2\,m_{1}\,p \Big]
   \nonumber \\ &+&
  {\phi}_{B}^{p}(x_{1})\,
  {\phi}_{\psi}^{V}(x_{2})\,
  {\phi}_{V}^{T}(x_{3})\, m_{1}\,m_{2}\,\Big[
  u\,(x_{3}-x_{1})+s\,(\bar{x}_{2}-x_{3})
  \Big] \Big\}\, {\alpha}_{s}(t_{c})
   \label{amp:ll-c-n},
   \end{eqnarray}
  %------------------------------------
  %------------------------------------
   \begin{eqnarray}
  {\cal A}^{LL}_{c,T} &=&
   \frac{1}{N_{c}}\,
  {\int}_{0}^{1}{\rm d}x_{1}
  {\int}_{0}^{1}{\rm d}x_{2}
  {\int}_{0}^{1}{\rm d}x_{3}
  {\int}_{0}^{\infty}{\rm d}b_{1}
  {\int}_{0}^{\infty}b_{2}\,{\rm d}b_{2}
  {\int}_{0}^{\infty}b_{3}\,{\rm d}b_{3}\,
  H_{n}({\alpha},{\beta}_{c},b_{2},b_{3})
   \nonumber \\ &{\times}&
  {\delta}(b_{1}-b_{3})\, \Big\{ {\phi}_{B}^{a}(x_{1})\,
  {\phi}_{\psi}^{T}(x_{2})\, m_{3}\,m_{c}\, \Big[
  2\,{\phi}_{V}^{V}(x_{3})
 -{\phi}_{V}^{A}(x_{3})\,t/(m_{1}\,p) \Big]
   \nonumber \\ &+&
  {\phi}_{B}^{p}(x_{1})\,
  {\phi}_{\psi}^{V}(x_{2})\,
  {\phi}_{V}^{T}(x_{3})\, 2\,m_{1}\,m_{2}\,
  (x_{1}-\bar{x}_{2}) \Big\}\, E_{n}(t_{c})\,
   {\alpha}_{s}(t_{c})
   \label{amp:ll-c-t},
   \end{eqnarray}
  %------------------------------------
  %-----------------------------------------------------
  %------------------------------------
   \begin{eqnarray}
  {\cal A}^{LR}_{c,P} &=&
   \frac{1}{N_{c}}\,
  {\int}_{0}^{1}{\rm d}x_{1}
  {\int}_{0}^{1}{\rm d}x_{2}
  {\int}_{0}^{1}{\rm d}x_{3}
  {\int}_{0}^{\infty}{\rm d}b_{1}
  {\int}_{0}^{\infty}b_{2}\,{\rm d}b_{2}
  {\int}_{0}^{\infty}b_{3}\,{\rm d}b_{3}\,
  {\delta}(b_{1}-b_{3})\,
  H_{n}({\alpha},{\beta}_{c},b_{2},b_{3})
   \nonumber \\ &{\times}&
   \Big\{
  {\phi}_{B}^{a}(x_{1})\, {\phi}_{P}^{a}(x_{3})\,
  2\, m_{1}\, p\, \Big[ {\phi}_{\psi}^{v}(x_{2})\,
  \{ u\,(x_{1}-x_{3})+t\,(x_{1}-\bar{x}_{2}) \}
  + {\phi}_{\psi}^{t}(x_{2})\, m_{2}\,m_{c} \Big]
  \nonumber \\ &+&
  {\phi}_{B}^{p}(x_{1})\, {\phi}_{\psi}^{v}(x_{2})\,
   m_{1}\,{\mu}_{P}\, \Big[ {\phi}_{P}^{p}(x_{3})\,
  2\,m_{1}\,p\,(x_{3}-x_{1})
  -{\phi}_{P}^{t}(x_{3})\, \{
  t\,(x_{1}-\bar{x}_{2})
  \nonumber \\ & &
   +s\,(\bar{x}_{2}-x_{3}) \} \Big]
  -{\phi}_{B}^{p}(x_{1})\, {\phi}_{\psi}^{t}(x_{2})\,
  {\phi}_{P}^{t}(x_{3})\, 4\,m_{1}\,m_{2}\,m_{c}\,{\mu}_{P}
    \Big\}\, E_{n}(t_{c})\, {\alpha}_{s}(t_{c})
   \label{amp:lr-c-p},
   \end{eqnarray}
  %------------------------------------
  %------------------------------------
   \begin{eqnarray}
  {\cal A}^{LR}_{c,L} &=&
   \frac{1}{N_{c}}\,
  {\int}_{0}^{1}{\rm d}x_{1}
  {\int}_{0}^{1}{\rm d}x_{2}
  {\int}_{0}^{1}{\rm d}x_{3}
  {\int}_{0}^{\infty}{\rm d}b_{1}
  {\int}_{0}^{\infty}b_{2}\,{\rm d}b_{2}
  {\int}_{0}^{\infty}b_{3}\,{\rm d}b_{3}\,
  {\delta}(b_{1}-b_{3})\,
  H_{n}({\alpha},{\beta}_{c},b_{2},b_{3})
   \nonumber \\ &{\times}&
   \Big\{ {\phi}_{B}^{a}(x_{1})\, {\phi}_{V}^{v}(x_{3})\,
   \Big[ {\phi}_{\psi}^{v}(x_{2})\,s\,\{
   t\,(\bar{x}_{2}-x_{1})+u\,(x_{3}-x_{1}) \}
 -{\phi}_{\psi}^{t}(x_{2})\,m_{2}\,m_{c}\,u \Big]
   \nonumber \\ &+&
  {\phi}_{B}^{p}(x_{1})\, {\phi}_{\psi}^{v}(x_{2})\,
  m_{1}\,m_{3}\, \Big[
  {\phi}_{V}^{s}(x_{3})\, 2\,m_{1}\,p\,(x_{1}-x_{3})
 + {\phi}_{V}^{t}(x_{3})\, \{ t\,(x_{1}-\bar{x}_{2})
  \nonumber \\ &+&
  s\,(\bar{x}_{2}-x_{3}) \}  \Big]
  + {\phi}_{B}^{p}(x_{1})\, {\phi}_{\psi}^{t}(x_{2})\,
  {\phi}_{V}^{t}(x_{3})\, m_{1}\,m_{3}\, (t-s)\, \frac{2\,m_{c}}{m_{2}}
   \Big\}\, E_{n}(t_{c})\, {\alpha}_{s}(t_{c})
   \label{amp:lr-c-l},
   \end{eqnarray}
  %------------------------------------
  %------------------------------------
   \begin{eqnarray}
  {\cal A}^{LR}_{c,N} &=&
   \frac{1}{N_{c}}\,
  {\int}_{0}^{1}{\rm d}x_{1}
  {\int}_{0}^{1}{\rm d}x_{2}
  {\int}_{0}^{1}{\rm d}x_{3}
  {\int}_{0}^{\infty}{\rm d}b_{1}
  {\int}_{0}^{\infty}b_{2}\,{\rm d}b_{2}
  {\int}_{0}^{\infty}b_{3}\,{\rm d}b_{3}\,
  {\delta}(b_{1}-b_{3})\,
  H_{n}({\alpha},{\beta}_{c},b_{2},b_{3})
   \nonumber \\ &{\times}&
   \Big\{ {\phi}_{B}^{p}(x_{1})\, {\phi}_{V}^{T}(x_{3})\,
   m_{1}\,\Big[ {\phi}_{\psi}^{V}(x_{2})\,m_{2}\,
   \{ u(x_{1}-x_{3})+s\,(x_{3}-\bar{x}_{2}) \}
 +{\phi}_{\psi}^{T}(x_{2})\,2\,m_{c}\,s \Big]
   \nonumber \\ &+&
   {\phi}_{B}^{a}(x_{1})\, \Big[
  {\phi}_{\psi}^{V}(x_{2})\, {\phi}_{V}^{V}(x_{3})\,
  2\,m_{2}\,m_{3}\, \{ t\,(\bar{x}_{2}-x_{1})
 +u\,(x_{3}-x_{1}) \}
  \nonumber \\ &-&
  {\phi}_{\psi}^{T}(x_{2})\,m_{3}\,m_{c} \{
  {\phi}_{V}^{V}(x_{3})\,t
 +{\phi}_{V}^{A}(x_{3})\,2\,m_{1}\,p \} \Big]
   \Big\}\, E_{n}(t_{c})\, {\alpha}_{s}(t_{c})
   \label{amp:lr-c-n},
   \end{eqnarray}
  %------------------------------------
  %------------------------------------
   \begin{eqnarray}
  {\cal A}^{LR}_{c,T} &=&
   \frac{1}{N_{c}}\,
  {\int}_{0}^{1}{\rm d}x_{1}
  {\int}_{0}^{1}{\rm d}x_{2}
  {\int}_{0}^{1}{\rm d}x_{3}
  {\int}_{0}^{\infty}{\rm d}b_{1}
  {\int}_{0}^{\infty}b_{2}\,{\rm d}b_{2}
  {\int}_{0}^{\infty}b_{3}\,{\rm d}b_{3}\,
  H_{n}({\alpha},{\beta}_{c},b_{2},b_{3})\, E_{n}(t_{c})
   \nonumber \\ &{\times}&
  {\delta}(b_{1}-b_{3})\,
   \Big\{  {\phi}_{B}^{p}(x_{1})\, {\phi}_{V}^{T}(x_{3})\,
   2\,m_{1}\, \Big[ {\phi}_{\psi}^{V}(x_{2})\,m_{2}\,
   (\bar{x}_{2}-x_{1})
  -{\phi}_{\psi}^{T}(x_{2})\,2\,m_{c} \Big]
   \nonumber \\ &+&
   {\phi}_{B}^{a}(x_{1})\, {\phi}_{\psi}^{V}(x_{2})\,
   {\phi}_{V}^{A}(x_{3})\, \frac{2\,m_{2}\,m_{3}}{m_{1}\,p}
   \{ u\,(x_{1}-x_{3})+ t\,(x_{1}-\bar{x}_{2}) \}
  \nonumber \\ &+&
  {\phi}_{B}^{a}(x_{1})\, {\phi}_{\psi}^{T}(x_{2})\,
  2\,m_{3}\,m_{c}\, \Big[ {\phi}_{V}^{V}(x_{3})
 +{\phi}_{V}^{A}(x_{3})\,\frac{t}{2\,m_{1}\,p} \Big]
   \Big\}\, {\alpha}_{s}(t_{c})
   \label{amp:lr-c-t},
   \end{eqnarray}
  %------------------------------------
  %-----------------------------------------------------
  %------------------------------------
   \begin{eqnarray}
  {\cal A}^{LL}_{d,P} &=&
   \frac{1}{N_{c}}\,
  {\int}_{0}^{1}{\rm d}x_{1}
  {\int}_{0}^{1}{\rm d}x_{2}
  {\int}_{0}^{1}{\rm d}x_{3}
  {\int}_{0}^{\infty}{\rm d}b_{1}
  {\int}_{0}^{\infty}b_{2}\,{\rm d}b_{2}
  {\int}_{0}^{\infty}b_{3}\,{\rm d}b_{3}\,
  {\delta}(b_{1}-b_{3})\,
  H_{n}({\alpha},{\beta}_{d},b_{2},b_{3})
   \nonumber \\ &{\times}&
   \Big\{
  {\phi}_{B}^{a}(x_{1})\, {\phi}_{P}^{a}(x_{3})\,
  2\, m_{1}\, p\, \Big[ {\phi}_{\psi}^{v}(x_{2})\,
  \{ u\,(x_{3}-x_{1})+t\,(x_{2}-x_{1}) \}
 -{\phi}_{\psi}^{t}(x_{2})\, m_{2}\,m_{c} \Big]
  \nonumber \\ &+&
  {\phi}_{B}^{p}(x_{1})\, {\phi}_{\psi}^{v}(x_{2})\,
   m_{1}\,{\mu}_{P}\, \Big[ {\phi}_{P}^{p}(x_{3})\,
  2\,m_{1}\,p\,(x_{1}-x_{3})
 +{\phi}_{P}^{t}(x_{3})\, \{ t\,(x_{1}-x_{2})
   \nonumber \\ &+&
  s\,(x_{2}-x_{3}) \} \Big]
 + {\phi}_{B}^{p}(x_{1})\, {\phi}_{\psi}^{t}(x_{2})\,
   {\phi}_{P}^{t}(x_{3})\, 4\, m_{1}\,m_{2}\,m_{c}\,{\mu}_{P}
    \Big\}\, E_{n}(t_{d})\, {\alpha}_{s}(t_{d})
   \label{amp:ll-d-p},
   \end{eqnarray}
  %------------------------------------
  %------------------------------------
   \begin{eqnarray}
  {\cal A}^{LL}_{d,L} &=&
   \frac{1}{N_{c}}\,
  {\int}_{0}^{1}{\rm d}x_{1}
  {\int}_{0}^{1}{\rm d}x_{2}
  {\int}_{0}^{1}{\rm d}x_{3}
  {\int}_{0}^{\infty}{\rm d}b_{1}
  {\int}_{0}^{\infty}b_{2}\,{\rm d}b_{2}
  {\int}_{0}^{\infty}b_{3}\,{\rm d}b_{3}\,
  {\delta}(b_{1}-b_{3})\,
  H_{n}({\alpha},{\beta}_{d},b_{2},b_{3})
   \nonumber \\ &{\times}&
   \Big\{ {\phi}_{B}^{a}(x_{1})\, {\phi}_{V}^{v}(x_{3})\,
   \Big[ {\phi}_{\psi}^{v}(x_{2})\,s\,
  \{ u\,(x_{1}-x_{3})+t\,(x_{1}-x_{2}) \}
 +{\phi}_{\psi}^{t}(x_{2})\, m_{2}\,m_{c}\,u \Big]
  \nonumber \\ &+&
  {\phi}_{B}^{p}(x_{1})\, {\phi}_{\psi}^{v}(x_{2})\,
   m_{1}\,m_{3}\, \Big[ {\phi}_{V}^{s}(x_{3})\,
  2\,m_{1}\,p\,(x_{3}-x_{1})
 +{\phi}_{V}^{t}(x_{3})\, \{ t\,(x_{2}-x_{1})
   \nonumber \\ &+&
  s\,(x_{3}-x_{2}) \} \Big]
 -{\phi}_{B}^{p}(x_{1})\, {\phi}_{\psi}^{t}(x_{2})\,
  {\phi}_{V}^{t}(x_{3})\, m_{1}\,m_{3}\,(t-s)\,\frac{2\,m_{c}}{m_{2}}
    \Big\}\, E_{n}(t_{d})\, {\alpha}_{s}(t_{d})
   \label{amp:ll-d-l},
   \end{eqnarray}
  %------------------------------------
  %------------------------------------
   \begin{eqnarray}
  {\cal A}^{LL}_{d,N} &=&
   \frac{1}{N_{c}}\,
  {\int}_{0}^{1}{\rm d}x_{1}
  {\int}_{0}^{1}{\rm d}x_{2}
  {\int}_{0}^{1}{\rm d}x_{3}
  {\int}_{0}^{\infty}{\rm d}b_{1}
  {\int}_{0}^{\infty}b_{2}\,{\rm d}b_{2}
  {\int}_{0}^{\infty}b_{3}\,{\rm d}b_{3}\,
  {\delta}(b_{1}-b_{3})\,
  H_{n}({\alpha},{\beta}_{d},b_{2},b_{3})
   \nonumber \\ &{\times}&
  E_{n}(t_{d})\, {\alpha}_{s}(t_{d})\,
   \Big\{ {\phi}_{B}^{a}(x_{1})\, m_{3}\, \Big[
  {\phi}_{\psi}^{V}(x_{2})\, {\phi}_{V}^{V}(x_{3})\,
  2\,m_{2}\, \{ u\,(x_{1}-x_{3})+t\,(x_{1}-x_{2}) \}
   \nonumber \\ &+&
  {\phi}_{\psi}^{T}(x_{2})\,m_{c}\, \{
  {\phi}_{V}^{V}(x_{3})\,t
 +{\phi}_{V}^{A}(x_{3})\,2\,m_{1}\,p \} \Big]
 -{\phi}_{B}^{p}(x_{1})\, {\phi}_{\psi}^{T}(x_{2})\,
  {\phi}_{V}^{T}(x_{3})\, 2\, m_{1}\,m_{c}\,s
   \nonumber \\ &+&
  {\phi}_{B}^{p}(x_{1})\, {\phi}_{\psi}^{V}(x_{2})\,
  {\phi}_{V}^{T}(x_{3})\, m_{1}\,m_{2}\,
  \{ u\,(x_{3}-x_{1})+s\,(x_{2}-x_{3}) \} \Big\}
   \label{amp:ll-d-n},
   \end{eqnarray}
  %------------------------------------
  %------------------------------------
   \begin{eqnarray}
  {\cal A}^{LL}_{d,T} &=&
   \frac{1}{N_{c}}\,
  {\int}_{0}^{1}{\rm d}x_{1}
  {\int}_{0}^{1}{\rm d}x_{2}
  {\int}_{0}^{1}{\rm d}x_{3}
  {\int}_{0}^{\infty}{\rm d}b_{1}
  {\int}_{0}^{\infty}b_{2}\,{\rm d}b_{2}
  {\int}_{0}^{\infty}b_{3}\,{\rm d}b_{3}\,
  {\delta}(b_{1}-b_{3})\,
  H_{n}({\alpha},{\beta}_{d},b_{2},b_{3})
   \nonumber \\ &{\times}&
  E_{n}(t_{d})\, {\alpha}_{s}(t_{d})\,
   \Big\{ {\phi}_{B}^{a}(x_{1})\, \Big[
   {\phi}_{\psi}^{V}(x_{2})\, {\phi}_{V}^{A}(x_{3})\,
    \frac{2\,m_{2}\,m_{3}}{m_{1}\,p}\,
    \{ u\,(x_{3}-x_{1})+t\,(x_{2}-x_{1}) \}
   \nonumber \\ &-&
  {\phi}_{\psi}^{T}(x_{2})\,m_{3}\,m_{c}\, \{
  2\,{\phi}_{V}^{V}(x_{3})+{\phi}_{V}^{A}(x_{3})
   \frac{t}{m_{1}\,p} \} \Big]
 +{\phi}_{B}^{p}(x_{1})\, {\phi}_{\psi}^{T}(x_{2})\,
  {\phi}_{V}^{T}(x_{3})\, 4\, m_{1}\, m_{c}
   \nonumber \\ &+&
  {\phi}_{B}^{p}(x_{1})\, {\phi}_{\psi}^{V}(x_{2})\,
  {\phi}_{V}^{T}(x_{3})\, 2\, m_{1}\, m_{2}\,
  (x_{1}-x_{2}) \Big\}
   \label{amp:ll-d-t},
   \end{eqnarray}
  %------------------------------------
  %-----------------------------------------------------
  %------------------------------------
   \begin{eqnarray}
  {\cal A}^{LR}_{d,P} &=&
   \frac{1}{N_{c}}\,
  {\int}_{0}^{1}{\rm d}x_{1}
  {\int}_{0}^{1}{\rm d}x_{2}
  {\int}_{0}^{1}{\rm d}x_{3}
  {\int}_{0}^{\infty}{\rm d}b_{1}
  {\int}_{0}^{\infty}b_{2}\,{\rm d}b_{2}
  {\int}_{0}^{\infty}b_{3}\,{\rm d}b_{3}\,
  H_{n}({\alpha},{\beta}_{d},b_{2},b_{3})
   \nonumber \\ &{\times}&
   E_{n}(t_{d})\, \Big\{
  {\phi}_{B}^{a}(x_{1})\, {\phi}_{P}^{a}(x_{3})\,
  2\, m_{1}\, p\, \Big[ {\phi}_{\psi}^{v}(x_{2})\,
  \{ u\,(x_{3}-x_{1})+s\,(x_{2}-x_{3}) \}
  \nonumber \\ &+&
  {\phi}_{\psi}^{t}(x_{2})\, m_{2}\,m_{c} \Big]
 +{\phi}_{B}^{p}(x_{1})\, {\phi}_{\psi}^{v}(x_{2})\,
   m_{1}\,{\mu}_{P}\, \Big[ {\phi}_{P}^{p}(x_{3})\,
  2\,m_{1}\,p\,(x_{1}-x_{3})
  \nonumber \\ &+&
 {\phi}_{P}^{t}(x_{3})\, \{ t\,(x_{2}-x_{1})
  +s\,(x_{3}-x_{2}) \} \Big] \Big\}\,
  {\alpha}_{s}(t_{d})\,{\delta}(b_{1}-b_{3})
   \label{amp:lr-d-p},
   \end{eqnarray}
  %------------------------------------
  %------------------------------------
   \begin{eqnarray}
  {\cal A}^{LR}_{d,L} &=&
   \frac{1}{N_{c}}\,
  {\int}_{0}^{1}{\rm d}x_{1}
  {\int}_{0}^{1}{\rm d}x_{2}
  {\int}_{0}^{1}{\rm d}x_{3}
  {\int}_{0}^{\infty}{\rm d}b_{1}
  {\int}_{0}^{\infty}b_{2}\,{\rm d}b_{2}
  {\int}_{0}^{\infty}b_{3}\,{\rm d}b_{3}\,
  H_{n}({\alpha},{\beta}_{d},b_{2},b_{3})
   \nonumber \\ &{\times}&
   E_{n}(t_{d})\, \Big\{
  {\phi}_{B}^{p}(x_{1})\, {\phi}_{\psi}^{v}(x_{2})\,
   m_{1}\,m_{3}\, \Big[ {\phi}_{V}^{t}(x_{3})\,
   \{ s\,(x_{2}-x_{3})+t\,(x_{1}-x_{2}) \}
   \nonumber \\ &+&
  {\phi}_{V}^{s}(x_{3})\,2\,m_{1}\,p\,(x_{3}-x_{1}) \Big]
 -{\phi}_{B}^{a}(x_{1})\,{\phi}_{V}^{v}(x_{3})\, \Big[
  {\phi}_{\psi}^{t}(x_{2})\, m_{2}\,m_{c}\,u
   \nonumber \\ &+&
  {\phi}_{\psi}^{v}(x_{2})\, 4\,m_{1}^{2}\,p^{2}\,(x_{2}-x_{1})
   \Big] \Big\}\,  {\alpha}_{s}(t_{d})\,{\delta}(b_{1}-b_{3})
   \label{amp:lr-d-l},
   \end{eqnarray}
  %------------------------------------
  %------------------------------------
   \begin{eqnarray}
  {\cal A}^{LR}_{d,N} &=&
   \frac{1}{N_{c}}\,
  {\int}_{0}^{1}{\rm d}x_{1}
  {\int}_{0}^{1}{\rm d}x_{2}
  {\int}_{0}^{1}{\rm d}x_{3}
  {\int}_{0}^{\infty}{\rm d}b_{1}
  {\int}_{0}^{\infty}b_{2}\,{\rm d}b_{2}
  {\int}_{0}^{\infty}b_{3}\,{\rm d}b_{3}\,
  H_{n}({\alpha},{\beta}_{d},b_{2},b_{3})
   \nonumber \\ &{\times}&
   E_{n}(t_{d})\, \Big\{
  {\phi}_{B}^{p}(x_{1})\, {\phi}_{\psi}^{V}(x_{2})\,
  {\phi}_{V}^{T}(x_{3})\, m_{1}\,m_{2}\,
   \{ u\,(x_{1}-x_{3})+s\,(x_{3}-x_{2}) \}
   \nonumber \\ &+&
  {\phi}_{B}^{a}(x_{1})\,{\phi}_{\psi}^{T}(x_{2})\,
   m_{3}\,m_{c}\, \Big[
   {\phi}_{V}^{A}(x_{3})\, 2\,m_{1}\,p
  -{\phi}_{V}^{V}(x_{3})\,t
   \Big] \Big\}\,  {\alpha}_{s}(t_{d})\,{\delta}(b_{1}-b_{3})
   \label{amp:lr-d-n},
   \end{eqnarray}
  %------------------------------------
  %------------------------------------
   \begin{eqnarray}
  {\cal A}^{LR}_{d,T} &=&
   \frac{1}{N_{c}}\,
  {\int}_{0}^{1}{\rm d}x_{1}
  {\int}_{0}^{1}{\rm d}x_{2}
  {\int}_{0}^{1}{\rm d}x_{3}
  {\int}_{0}^{\infty}{\rm d}b_{1}
  {\int}_{0}^{\infty}b_{2}\,{\rm d}b_{2}
  {\int}_{0}^{\infty}b_{3}\,{\rm d}b_{3}\,
  H_{n}({\alpha},{\beta}_{d},b_{2},b_{3})
   \nonumber \\ &{\times}&
   E_{n}(t_{d})\, {\alpha}_{s}(t_{d})\, \Big\{
  {\phi}_{B}^{a}(x_{1})\, {\phi}_{\psi}^{T}(x_{2})\,
  m_{3}\,m_{c}\, \Big[
   {\phi}_{V}^{A}(x_{3})\, \frac{t}{m_{1}\,p}
  -2\,{\phi}_{V}^{V}(x_{3}) \Big]
  \nonumber \\ &+&
  {\phi}_{B}^{p}(x_{1})\, {\phi}_{\psi}^{V}(x_{2})\,
  {\phi}_{V}^{T}(x_{3})\, 2\,m_{1}\,m_{2}\,(x_{2}-x_{1})
   \Big] \Big\}\,  {\delta}(b_{1}-b_{3})
   \label{amp:lr-d-t},
   \end{eqnarray}
  %------------------------------------
  %-----------------------------------------------------
  where $x_{i}$ and $b_{i}$ are the longitudinal momentum fraction
  and the conjugate variable of the transverse momentum $k_{iT}$,
  respectively. The subscript $i$ of ${\cal A}_{i,j}^{k}$ corresponds
  to the indices of Fig.\ref{fig:fey}; the subscript $j$ $=$ $P$,
  $L$, $N$, $T$ correspond to the different helicity amplitudes;
  the superscript $k$ refers to the two possible Dirac structures
  ${\Gamma}_{1}{\otimes}{\Gamma}_{2}$ of the operators
  $(\bar{q}_{1}q_{2})_{{\Gamma}_{1}}(\bar{q}_{3}q_{4})_{{\Gamma}_{2}}$,
  namely $k$ $=$ $LL$ for $(V-A){\otimes}(V-A)$ and $k$ $=$ $LR$
  for $(V-A){\otimes}(V+A)$.

  The function $H_{f,n}$ and Sudakov factor $E_{f,n}$ are
  defined as
  %-----------------------------------------------------
   \begin{equation}
   H_{f}({\alpha},{\beta},b_{i},b_{j})\, =\,
   K_{0}(b_{i}\sqrt{-{\alpha}})\, \Big\{
  {\theta}(b_{i}-b_{i}) K_{0}(b_{i}\sqrt{-{\beta}})\,
   I_{0}(b_{j}\sqrt{-{\beta}})
   + (b_{i} {\leftrightarrow} b_{j}) \Big\}
   \label{amp:hab},
   \end{equation}
  %-----------------------------------------------------
   \begin{eqnarray}
  H_{n}({\alpha},{\beta},b_{i},b_{j}) &=&
  \Big\{ {\theta}(-{\beta})\, K_{0}(b_{i}\sqrt{-{\beta}})
  +\frac{{\pi}}{2}\,
  {\theta}({\beta})\, \Big[ i\,J_{0}(b_{i}\sqrt{{\beta}})
   - Y_{0}(b_{i}\sqrt{{\beta}}) \Big] \Big\}
   \nonumber \\ &{\times}&
   \Big\{ {\theta}(b_{i}-b_{j})\, K_{0}(b_{j}\sqrt{-{\alpha}})\,
   I_{0}(b_{j}\sqrt{-{\alpha}}) + (b_{i} {\leftrightarrow} b_{j}) \Big\}
   \label{amp:hcd},
   \end{eqnarray}
  %-----------------------------------------------------
   \begin{equation}
   E_{f}(t)\ =\ {\exp}\{ -S_{B}(t)-S_{M}(t) \}
   \label{sudakov-f},
   \end{equation}
  %-----------------------------------------------------
   \begin{equation}
   E_{n}(t)\ =\ {\exp}\{ -S_{B}(t)-S_{M}(t)-S_{\psi}(t) \}
   \label{sudakov-n},
   \end{equation}
  %-----------------------------------------------------
   \begin{equation}
  S_{B}(t)\, =\, s(x_{1},b_{1},p_{1}^{+})
  +2{\int}_{1/b_{1}}^{t}\frac{d{\mu}}{\mu}{\gamma}_{q}
   \label{sudakov-bu},
   \end{equation}
  %-----------------------------------------------------
   \begin{equation}
  S_{M}(t)\, =\, s(x_{3},b_{3},p_{3}^{+}) + s(\bar{x}_{3},b_{3},p_{3}^{+})
  +2{\int}_{1/b_{3}}^{t}\frac{d{\mu}}{\mu}{\gamma}_{q}
   \label{sudakov-qq},
   \end{equation}
  %-----------------------------------------------------
   \begin{equation}
  S_{\psi}(t)\, =\, s(x_{2},b_{2},p_{2}^{+}) + s(\bar{x}_{2},b_{2},p_{2}^{+})
  +2{\int}_{1/b_{2}}^{t}\frac{d{\mu}}{\mu}{\gamma}_{q}
   \label{sudakov-cc},
   \end{equation}
  %-----------------------------------------------------
  where $I_{0}$, $J_{0}$, $K_{0}$ and $Y_{0}$ are Bessel
  functions; ${\gamma}_{q}$ $=$ $-{\alpha}_{s}/{\pi}$ is the
  quark anomalous dimension; the expression of $s(x,b,Q)$
  can be found in the appendix of Ref.\cite{prd52.3958};
  ${\alpha}$ and ${\beta}$ are the virtualities of gluon
  and quarks. The subscript of the quark virtuality ${\beta}_{i}$
  corresponds to the indices of Fig.\ref{fig:fey}.
  The definitions of the particle virtuality and typical
  scale $t_{i}$ are given as follows.
  %-----------------------------------------------------
   \begin{eqnarray}
  {\alpha} &=& x_{1}^{2}\,m_{1}^{2}+x_{3}^{2}\,m_{3}^{2}-x_{1}\,x_{3}\,u
   \label{alpha-gluon}, \\
  %-----------------------------------------------------
  {\beta}_{a} &=& x_{3}^{2}\,m_{3}^{2}-x_{3}\,u+m_{1}^{2}-m_{b}^{2}
   \label{beta-a}, \\
  %-----------------------------------------------------
  {\beta}_{b} &=& x_{1}^{2}\,m_{1}^{2}-x_{1}\,u+m_{3}^{2}
   \label{beta-b}, \\
  %-----------------------------------------------------
  {\beta}_{c} &=& {\alpha}+\bar{x}_{2}^{2}\,m_{2}^{2}
       -x_{1}\,\bar{x}_{2}\,t+\bar{x}_{2}\,x_{3}\,s-m_{c}^{2}
   \label{beta-c}, \\
  %-----------------------------------------------------
  {\beta}_{d} &=& {\alpha}+x_{2}^{2}\,m_{2}^{2}
       -x_{1}\,x_{2}\,t+x_{2}\,x_{3}\,s-m_{c}^{2}
   \label{beta-d}, \\
   t_{i} &=&
  {\max}(\sqrt{{\vert}{\alpha}{\vert}},\sqrt{{\vert}{\beta}_{i}{\vert}},1/b_{1},1/b_{2},1/b_{3})
   \label{scale-ti}.
   \end{eqnarray}
  %-----------------------------------------------------
  \end{appendix}

  %%%%%%%%%%%%%%%%%%%%%%%%%%%%%%%%%%%%%%%%%%%%%%%%%%%%%%%%%%%
  
  \end{document}